\documentclass[aps,prb,twocolumn,10pt]{revtex4-1}
\usepackage[english]{babel}
\usepackage[utf8]{inputenc}

\usepackage{graphicx}
\usepackage{xcolor}
\usepackage{tabularx}
\usepackage{booktabs}
\usepackage{amssymb}
\usepackage{amsmath}
\usepackage{upgreek}
\usepackage{multirow}
\usepackage{silence}

\newcommand{\w}{\omega}
\newcommand{\ep}{\epsilon}
\newcommand{\imag}{i}
\newcommand{\E}{\mathbf{E}}
\newcommand{\B}{\mathbf{B}}
\newcommand{\J}{\mathbf{J}}

\newcommand{\mob}{\mathrm{cm^2 V^{-1} s^{-1}}}
\newcommand{\cmm}{\mathrm{cm^{-3}}}
\newcommand{\tss}[1]{\textsuperscript{#1}}

\newcommand{\figwid}{\columnwidth}
\newcommand{\etal}{\emph{et al.}}
\newcommand{\citen}[1]{[\onlinecite{#1}]}

\begin{document}

\title{Size-dependent nonlocal effects in plasmonic semiconductor particles}
\author{Johan~R.~Maack$^1$} 
\author{N.~Asger~Mortensen$^{2,3}$}
\author{Martijn~Wubs$^{1,3}$}
\affiliation{$^1$Department of Photonics Engineering, Technical University of Denmark, \O rsteds Plads 343, DK-2800~Kongens~Lyngby, Denmark\\
$^2$Center for Nano Optics, University of Southern Denmark, Campusvej 55, DK-5230~Odense~M, Denmark\\
$^3$Center for Nanostructured Graphene, Technical University of Denmark, \O rsteds Plads 343, DK-2800 Kongens Lyngby, Denmark}
\date{\today}

\begin{abstract}
Localized surface plasmons (LSP) in semiconductor particles are expected to exhibit spatial nonlocal response effects as the geometry enters the nanometer scale. To investigate these nonlocal effects, we apply the hydrodynamic model to nanospheres of two different semiconductor materials: intrinsic InSb and $n$-doped GaAs. Our results show that the semiconductors indeed display nonlocal effects, and that these effects are even more pronounced than in metals. In a $150\mathrm{\,nm}$ InSb particle at $300\mathrm{\,K}$, the LSP frequency is blueshifted 35\%, which is orders of magnitude larger than the blueshift in a metal particle of the same size. This property, together with their tunability, makes semiconductors a promising platform for experiments in nonlocal effects.
\end{abstract}

\maketitle

\section{Introduction}
It has been known for a while that the Drude model for metals is only applicable when the geometry is sufficiently large compared to intrinsic length scales of the electron gas. When analyzing nanoscale structures, the model becomes less accurate and a different or augmented model becomes necessary. A model which has successfully described metal structures on the nanoscale is the hydrodynamic Drude model (HDM)\cite{abajo08,mcmahon09, raza13,toscano15,christensen14}, where wavevector dependence is added to the Drude dielectric function. Due to this, the model has been able to explain observable nonlocal effects, such as longitudinal waves inside the metal and a size-dependent shift of the resonance frequency of localized surface plasmons (LSP)\cite{raza15}. 

However, the HDM is not necessarily restricted to metals, but could be relevant for other nanoscale structures with a free electron-like plasma as well. In this paper, we consider the application of the HDM to semiconductors, where the charge carriers are electrons and/or holes. This leads to new predictions, different from the well known insights obtained by application of the usual Drude model to semiconductors\cite{Fox}. Among the most notable differences between metals and semiconductors are the densities and the effective masses of the charge carriers. Metals have large free carrier concentrations and effective masses roughly equal to that of the free electron. Semiconductors on the other hand mostly have lower charge carrier densities, and these furthermore depend strongly on doping level and temperature. The effective masses will vary from material to material, and usually the effective masses of holes and electrons are different. 

As briefly mentioned by Hanham \etal\cite{hanham12}, these characteristics can be exploited to investigate nonlocal effects in ways that are not immediately possible in metals. By using semiconductors, the frequency of operation shifts from the optical spectrum to the infrared or THz bands because the plasma frequency, which depends on the charge carrier density, is lower than in metals. As we predict here, the size-dependent nonlocal effects will simultaneously manifest themselves in larger structures than in metals, which is good news for both fabrication and observation. 

The optical properties of semiconductors have already been described by many semiclassical and quantum mechanical models (see for example \citen{haug}). In particular, semiconductors are known to exhibit quantum confinement when the size of the structure is on the nanometer scale, such as in quantum wells and dots\cite{kelly}. But in some cases, the plasma description is more suitable. An example is InSb which is characterized by an extremely small band gap ($E_g\sim\! 0.17\mathrm{\,eV}$) and a high charge carrier density at room temperature. This material was used by Hanham \etal, as well as in earlier papers on plasmonics\cite{ritz84,jones95,rivas06}, and in all cases the charge carriers were treated as a plasma. Another example is doped semiconductors where additional charge carriers have been supplied by the donors or acceptors. Plasmonics in doped semiconductors has additional advantages such as tunability\cite{naik13}, and plasmonic experiments with both $n$- and $p$-doping have been conducted\cite{matz81,grychowski86,betti89,chen89,meng91,biagi92,ginn11}. 

In the region between semiconductors described by a plasma model (such as the Drude model) and quantum dots is a transition-zone, where neither macroscopic nor microscopic theories are ideal. This region, which is defined by the size of the structures as well as the number of charge carriers, has been the subject of both experimental\cite{luther11,garcia11,buonsanti11,manthiram12,schimpf13,zhou15} and theoretical\cite{hapala13,zhang14,faucheaux14,carmina15,zhang17} studies. In this paper we will investigate semiconductor particles that are large enough to contain sufficient charge carriers to be described by a plasma model, yet small enough to display nonlocal effects (implying that the Drude model becomes inaccurate). We will focus on spherical particles of intrinsic InSb and $n$-doped GaAs and use the HDM to calculate the optical properties. To set a lower limit of our model, we use the results from Zhang\etal\cite{zhang14} who estimated the onset of quantum confinement effects in semiconductors using first-principles calculations. Although they find no hard transition, their results show that for a nanoparticle with a radius of $2.5\mathrm{\,nm}$ and only a few charge carriers, the plasma model is able to reproduce the DFT calculations reasonably well. But to make sure we are in the plasma regime, we will only consider particles containing more than 50 charge carriers (and, as seen in the results section,  radii much larger than $2.5\mathrm{\,nm}$).

We will mainly look at intraband transitions, as these affect the properties of the plasma directly, while interband transitions for simplicity are ignored. This is a reasonable assumption as long as the energies considered are smaller than the band gap. Another kind of excitations characteristic of semiconductors is excitons, which give rise to energy levels inside the band gap and modifications to the conduction band edge. However, for materials with a very narrow band gap, like InSb, the excitons are bound so weakly that they usually can be neglected\cite{haug}. Similarly for doped semiconductors, the screening effect of the high charge carrier density weakens the excitonic bond. For the materials that we study here, it is therefore a reasonable approximation to ignore exciton effects.

Given the assumptions above, the hydrodynamic equations of motion can be rederived for charge carriers in semiconductors, and in the next section, the key expressions in the model will be presented. These expressions will then be applied to spherical nanoparticles, and finally the results of the numerical simulations will be discussed.

\section{The model}
\label{model}
\subsection{Dielectric functions}
The hydrodynamic Drude model is characterized by a nonlocal longitudinal dielectric function\cite{wubs15,Raza13a}
\begin{equation}
\label{eq:longe}
\ep_L(k,\w)=\ep_{\infty}-\frac{\w_p^2}{\w^2+\imag\gamma\w-\beta^2 k^2} \:,
\end{equation}
where $\w_p$ is the plasma frequency, $\gamma$ is the damping rate, $\ep_{\infty}$ is the background dielectric constant, and $\beta$ is a parameter that describes the strength of nonlocality. In this paper, $\ep_{\infty}$ is chosen to be constant in $\w$, which is a good approximation for energies smaller than the band gap.

For the degenerate electron gas in metals, $\beta$ is directly related to the Fermi velocity $v_F$ (see Refs. \citen{christensen14,raza15}), but for semiconductors, the parameter depends on several conditions. The most obvious complication in semiconductors compared to metals is the presence of more than one kind of charge carrier, including electrons and heavy and light holes. The electrons, however, have a much smaller effective mass than the holes for a typical semiconductor, and therefore they will determine the optical properties almost entirely. This means that the holes can be ignored as a first approximation whenever electrons are present as majority charge carriers, as they are in this paper. Semiconductors also differ from metals in the sense that changes in charge carrier densities can be created by different means. If the electrons are thermally excited to the conduction band, and the bands are assumed to be parabolic, one can derive the expression for the dielectric function using a simple quantum mechanical model similar to the Lindhard model (see Supporting Information). In this derivation, $\beta$ is given by
\begin{equation}
\label{eq:beta_therm}
\beta^2=\frac{3 k_B T}{m_e^*} \:,
\end{equation}
where $m_e^*$ is the effective mass of the electron, $T$ is the temperature, and $k_B$ is the Boltzmann constant. This expression is only valid for low temperatures where the Fermi--Dirac distribution can be approximated with the Boltzmann distribution. If this is not the case, the value of $\beta$ can be found with numerical methods. 

If the semiconductor instead is $n$-doped (and we neglect electrons thermally excited from the valence band to the conduction band), then $\beta$ is given by
\begin{equation}
\label{eq:beta_dope}
\beta^2=\frac{3}{5}v_F^2=\frac{3}{5}\frac{\hbar^2}{{m_e^*}^2}\left(3 \pi^2 n\right)^{\frac{2}{3}} ,
\end{equation}
where $n$ is the electron density. Equation (\ref{eq:beta_dope}) can also be used if the charge carriers are created by an external energy source, \emph{e.g.} a laser pulse that can excite carriers across the band gap. This situation would of course be complicated by the relaxation of the charge carriers over time, and assumptions about a quasi-equilibrium would have to be made (and we will not consider this here). Note, that the two expressions for $\beta$ also can be found in \citen{jones95}.

The equations for the plasma frequency $\w_p$ and the damping rate $\gamma$, however, are independent of the excitation method and in all cases are given by
\begin{align}
\label{eq:plasma}
\w_{p}^2=\frac{n e^2}{\ep_0 m_e^*} \;, \\
\label{eq:gamma}
\gamma=\frac{e}{m_{e,cond}^*\mu_e} \;,
\end{align}
where $\mu_e$ is the mobility of the electron. Here $m_{e,cond}^*$ is the \emph{conductivity} effective mass of the electron, and this is in general different from $m_e^*$ (which is called the \emph{density-of-states} effective mass). Only for isotropic and perfectly parabolic bands are they identical\cite{sze}.

For doped semiconductors, $n$ is equal to the doping concentration $N_d$ if the donors are completely ionized (which is a good approximation at room temperature). For thermally excited electrons in intrinsic semiconductors, $n$ is given by\cite{sze}
\begin{equation}
\label{eq:n_therm}
n=2\left(\frac{2\pi k_B T}{h^2}\right)^{\frac{3}{2}} m_e^{*\frac{3}{4}} m_h^{*\frac{3}{4}}\exp\left(-\frac{E_g}{2 k_B T}\right) ,
\end{equation}
\noindent where $m_{h}^*$ is the density-of-states effective mass of the holes. The equation is valid when the Boltzmann distribution is accurate, but numerical methods can be used to find $n$ if this is not the case. 

While the longitudinal dielectric function  in Eq. (\ref{eq:longe}) is nonlocal in the HDM, the transversal dielectric function is local\cite{wubs15,Raza13a}, \emph{i.e.}
\begin{equation}
\label{eq:transe}
\ep_T(\w)=\ep_{\infty}-\frac{\w_p^2}{\w^2+\imag\gamma\w} \:.
\end{equation}

\subsection{The hydrodynamic equations}
The two dielectric functions together with Maxwell's equations produce the following equations in real space\cite{raza11,toscano12,raza15}
\begin{subequations}
\begin{align}
\label{eq:wave}
-\nabla\times\nabla\times\E+ \frac{\w^2}{c^2}\ep_{\infty}\E=-\imag\mu_0\w\J \,, \\
\label{eq:hydro}
\frac{\beta^2}{\w^2+\imag\gamma\w}\nabla\left(\nabla\cdot\J\right)+ \J=\frac{\imag\w\ep_0\w_{p}^2}{\w^2+\imag\gamma\w}\E \:,
\end{align}
\end{subequations}
where the first is the classical wave equation, and the second is the linearized nonlocal hydrodynamic equation. These equations provide a relation between the electrical field $\E$ and the induced current density $\J$. In a local approximation ($\beta\approx 0$), Eq. \ref{eq:hydro} would reduce to Ohm's law, \emph{i.e.} $\J\propto\E$ with the constant of proportionality given by the usual Drude conductivity $\sigma_D=\imag\w\ep_0\w_{p}^2/(\w^2+\imag\gamma\w)$. The relation between these equations and  $\ep_L(k,\w)$ and $\ep_T(\w)$ is easily seen for an infinite medium by using a Fourier transform\cite{wubs15,raza15}.

Equations (\ref{eq:wave}) and (\ref{eq:hydro}) can be solved for various geometries when provided with the necessary boundary conditions, using either analytical approaches or numerical methods\cite{david11,toscano12,toscano13,yan13,Raza13a,christensen14}. The continuities of $\E_{\parallel}$ and $\B_{\parallel}$ across the boundary are the natural first two boundary conditions. However, an additional third boundary condition is needed in the case of the HDM. Under the assumption of an infinite work function, the boundary condition is $\J_{\perp}=\mathbf{0}$, \emph{i.e.} the charge carriers cannot escape the material (see \citen{raza15} for a discussion). This choice implies that the spill-out of electrons at the interface is ignored\cite{toscano15}.

\subsection{The Mie coefficients}
Given the boundary conditions, the solutions for $\E$ and $\J$ are found for spherical symmetry. This was originally done by Mie for transversal waves\cite{mie08}, and then later Ruppin added the longitudinal component which is present for the HDM\cite{ruppin73}. The final result is contained in two transversal coefficients denoted by $a_n^j$ and $b_n^j$ and one longitudinal coefficient denoted by $c_n^j$. Here $n$ is an integer, and $j$ indicates whether the field is reflected from the sphere ($j=r$) or transmitted into the sphere ($j=t$). The coefficient $c_n^r$ is zero as the surrounding medium is assumed to be a dielectric and unable to support longitudinal waves. 

However, because our additional boundary condition is different from Ruppin's, we will instead of his results use the solution from David \etal\cite{david11} where the reflection coefficients are given by
\begin{subequations}
\begin{align}
\label{eq:mie_a}
&a_n^r=\frac{-j_n(x_D)\lbrack x_T j_n(x_T)\rbrack^\prime
             +j_n(x_T)\lbrack x_D j_n(x_D)\rbrack^\prime}
      {h^{(1)}_n(x_D)\lbrack x_T j_n(x_T)      \rbrack^\prime
      -j_n(x_T)      \lbrack x_D h^{(1)}_n(x_D)\rbrack^\prime} \:, \\
\label{eq:mie_b}
&b_n^r= \nonumber \\
& \frac{-\ep_D j_n(x_D)\left(\Delta_n\!+\!\lbrack x_T j_n(x_T)\rbrack^\prime \right)
                             \!+\!\ep_T j_n(x_T)\lbrack x_D j_n(x_D)\rbrack^\prime}
           {\ep_D h_n^{(1)}(x_D)\left(\Delta_n\!+\!\lbrack x_T j_n(x_T)\rbrack^\prime \right)
                      \!-\!\ep_T j_n(x_T)\lbrack x_D h_n^{(1)}(x_D)\rbrack^\prime} \:.
\end{align}
\end{subequations}
Here, $x_D=R k_D=\sqrt{\ep_D} R\w/c$ and $x_T=R k_T=\sqrt{\ep_T} R\w/c$. The parameter $\ep_D$ is the dielectric constant of the surrounding dielectric and $\ep_T$ is given by Eq. (\ref{eq:transe}). The function $j_n$ is the spherical Bessel function of the first kind, and $h_n^{(1)}$ is the spherical Hankel function of the first kind. The differentiation (denoted with the prime) is with respect to the argument. The nonlocal parameter $\Delta_n$ is given by\cite{david11}
\begin{equation}
\label{eq:mie_d}
\Delta_n=\frac{j_n(x_T) j_n(x_L) n(n+1)}{x_L j_n^\prime(x_L)}\left(\frac{\ep_T}{\ep_{\infty}}-1\right) ,
\end{equation}
where $x_L=R k_L$ and the longitudinal wave vector is\cite{david11}
\begin{equation}
\label{eq:longk}
k_L=\frac{1}{\beta}\sqrt{\w^2+\imag\gamma\w-\frac{\w_p^2}{\ep_{\infty}}} \:.
\end{equation}
The coefficients $a_n$ and $b_n$ are related to oscillations of the magnetic and electric type, respectively. Note that the expression for $a_n$ is identical to the classical local solution, while the expression for $b_n$ is not\cite{bohren}. Setting $\Delta_n=0$, however, reduces the $b_n$ coefficients to their classical local-response counterparts as well. 

Once the $a_n^r$ and $b_n^r$ coefficients are known, the extinction cross-section for single particles can be found with\cite{bohren}
\begin{equation}
\label{eq:sigma}
\sigma_{ext}=-\frac{2 \pi}{k_D^2}\sum_{n=1}(2 n+1)\mathrm{Re}(a_n^r+b_n^r) \:.
\end{equation}

\section{Results}
\label{results}

Using Eq. (\ref{eq:sigma}), the extinction spectra for spherical semiconductor nanoparticles will now be determined. To begin with, we will look at intrinsic InSb with thermally excited charge carriers. The data for InSb at $T=300\mathrm{\,K}$ is given in table \ref{tab:data}, and using equations (\ref{eq:beta_therm}), (\ref{eq:plasma}) and (\ref{eq:gamma}) we find $\beta=1.07\times 10^{6}\mathrm{\,m/s}$, $\w_p=6.94\times 10^{13}\mathrm{\,s^{-1}}$ and $\gamma=1.94\times 10^{12}\mathrm{\,s^{-1}}$. From the plasma frequency it is immediately seen that excitation of the plasmon must take place in the infrared domain.  

\begin{table}
\caption{Properties of GaAs and InSb. The intrinsic charge carrier density is denoted by $n_i$. The masses $m_e^*$ and $m_h^*$ for InSb are taken from \citen{stradling70} and \citen{cunningham70} respectively. For GaAs, $m_e^*$ and $m_{e,cond}^*$ (which depends on the doping level $N_d$) are from \citen{szmyd90}, and $m_h^*$ is from \citen{walton68}. $E_g$ for InSb is taken from \citen{rowell88}, and $\mu_e$ and $\mu_h$ for GaAs are from \citen{sze67}. The rest of the data are taken from \citen{madelung}. Note that for InSb, the conductivity effective mass is assumed to be identical to the density-of-states effective mass.}
\setlength\tabcolsep{2.5pt} 
\begin{tabular}{lrrrr}
\toprule[0.5pt]
                       & \multicolumn{2}{r}{GaAs ($300\mathrm{\,K}$)} & InSb ($300\mathrm{\,K}$)    & InSb ($200\mathrm{\,K}$)    \\
\midrule[0.5pt]
$\ep_{\infty}$                      & & $ 10.9 $                     & $ 15.7 $                    & $ 15.7 $                     \\
$E_g$ (eV)                          & & $ 1.42 $                     & $ 0.17 $                    & $ 0.20 $                     \\
$n_i$ ($\cmm$)                      & & $ 2.1\times 10^6 $           & $ 1.9\times 10^{16} $       & $ 8.6\times 10^{14} $        \\
\multicolumn{2}{l}{\multirow{2}{*}{$\mu_e$ ($\mob$)}} & $ 2900 $\tss{a} & \multirow{2}{*}{$ 77000 $} & \multirow{2}{*}{$ 151000 $} \\
\multicolumn{2}{l}{}                & $ 1100 $\tss{b}                 &                             &                              \\
\multicolumn{2}{l}{\multirow{2}{*}{$\mu_h$ ($\mob$)}} &  $ 190 $\tss{a} & \multirow{2}{*}{$ 850 $}   & \multirow{2}{*}{$ 1910 $}  \\
\multicolumn{2}{l}{}                & $ 80 $\tss{b}                   &                             &                              \\
$m_e^*/m_0$                         & & $ 0.0636 $                    & $ 0.0118 $                  & $ 0.0126 $                   \\
$m_h^*/m_0$                         & & $ 0.53 $                      & $ 0.48 $                    & $ 0.44 $                     \\
\multirow{2}{*}{$m_{e,cond}^*/m_0$} & & $ 0.0695 $\tss{a}             & \multirow{2}{*}{$ 0.0118 $} & \multirow{2}{*}{$ 0.0126 $}  \\
                                    & & $ 0.101 $\tss{b}              &                             &                               \\
\bottomrule[0.5pt]
\multicolumn{5}{l}{\tss{a}\footnotesize{$N_d=10^{18}$ $\cmm$}} \\
\multicolumn{5}{l}{\tss{b}\footnotesize{$N_d=10^{19}$ $\cmm$}}
\end{tabular}
\label{tab:data}
\end{table}

\begin{figure}
    \includegraphics[width=\figwid]{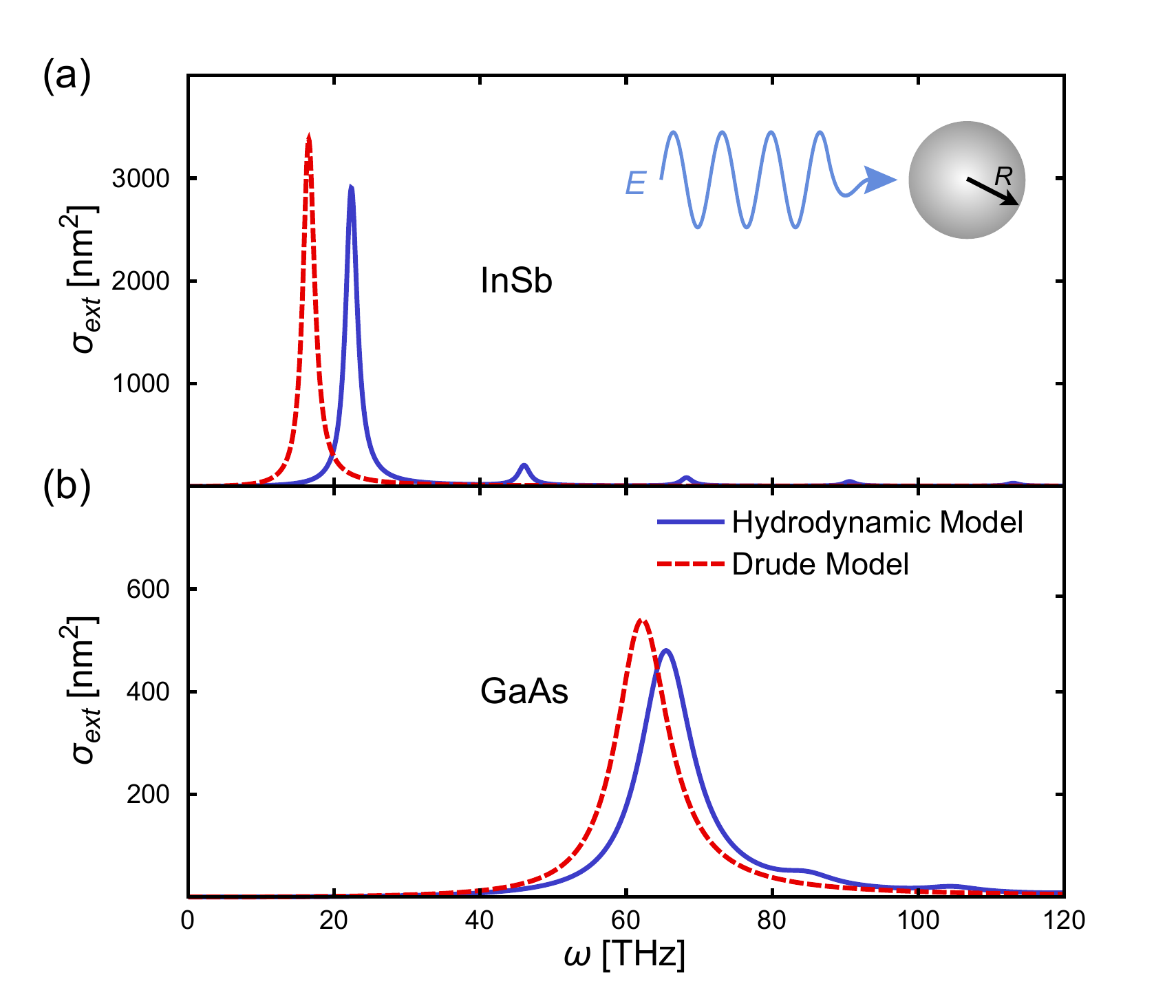}
\caption{\label{fig:insb_gaas}(a) Extinction spectrum for an InSb nanoparticle in vacuum with $R=150\mathrm{\,nm}$. Charge carriers are thermally excited, and the temperature is $300\mathrm{\,K}$. (b) Extinction spectrum for a GaAs nanoparticle in vacuum with $R=50\mathrm{\,nm}$. The doping level is $N_d=10^{18}\mathrm{\,cm^{-3}}$. The dashed line is the local Drude model, and the full line is the HDM. Material parameters can be found in table \ref{tab:data}.}
\end{figure}

In Fig. \ref{fig:insb_gaas}(a), the extinction cross section for an InSb nanoparticle at $T=300\mathrm{\,K}$ in vacuum with $R=150\mathrm{\,nm}$ is plotted. The dashed line is the local-response approximation obtained by setting $\Delta_n$ equal to zero in Eq. (\ref{eq:mie_b}). This curve only has a single visible peak which can be recognized as the classical dipole plasmon peak with a frequency close to $\w_{dipole}=\w_p/(\ep_{\infty}+2\ep_D)^{1/2}$. Peaks from higher-order poles also exist, but are too faint to see here. The full line in the figure is the hydrodynamic solution, and it differs from the classical local-response result in several ways. The first thing we notice is that the dipole peak is shifted towards higher frequencies, and secondly we see that new peaks above the plasma frequency have appeared. The new peaks and the blueshift are clear signatures of nonlocality, and are well known phenomena in metals\cite{tiggesbaumker93,lindau71,raza11,christensen14,raza15}. There, the peaks are known to be associated with confined bulk plasmons, and the blueshift of the dipole peak is found to increase as the particle gets smaller\cite{christensen14}. The existence of such nonlocal effects in semiconductors has, to our knowledge, not been predicted before. Furthermore, the blueshift in Fig. \ref{fig:insb_gaas}(a) is significant, thus facilitating the experimental verification by, for instance, systematically measuring the peak position as a function of particle size.

By using doping, wide-gap semiconductors can also be used as plasmonic materials. To investigate the predictions of the HDM for doped semiconductors, we will consider $n$-doped GaAs with a donor (\emph{e.g.} silicon\cite{matz81}) concentration of $N_d=10^{18}\mathrm{\,cm^{-3}}$. The data for GaAs is shown in table \ref{tab:data}, and using equations (\ref{eq:beta_dope}), (\ref{eq:plasma}) and (\ref{eq:gamma}) we find $\beta=4.36\times 10^{5}\mathrm{\,m/s}$, $\w_p=2.24\times 10^{14}\mathrm{\,s^{-1}}$ and $\gamma=8.72\times 10^{12}\mathrm{\,s^{-1}}$. In Fig. \ref{fig:insb_gaas}(b) the extinction spectrum for a doped GaAs nanoparticle with $R=50\mathrm{\,nm}$ is plotted. Once again we see oscillations above the plasma frequency and a clear blueshift. 

Although the results in Fig. \ref{fig:insb_gaas}(a) and \ref{fig:insb_gaas}(b) appear promising, it should be noted that the amplitudes of the signals are about a hundred times weaker than the signal from, for example, a silver particle of the same size. Experimental sensitivity is improving, however, and at least one group has already measured signals of the same magnitude as the ones predicted here\cite{billaud10}. 

\begin{figure}
    \includegraphics[width=\figwid]{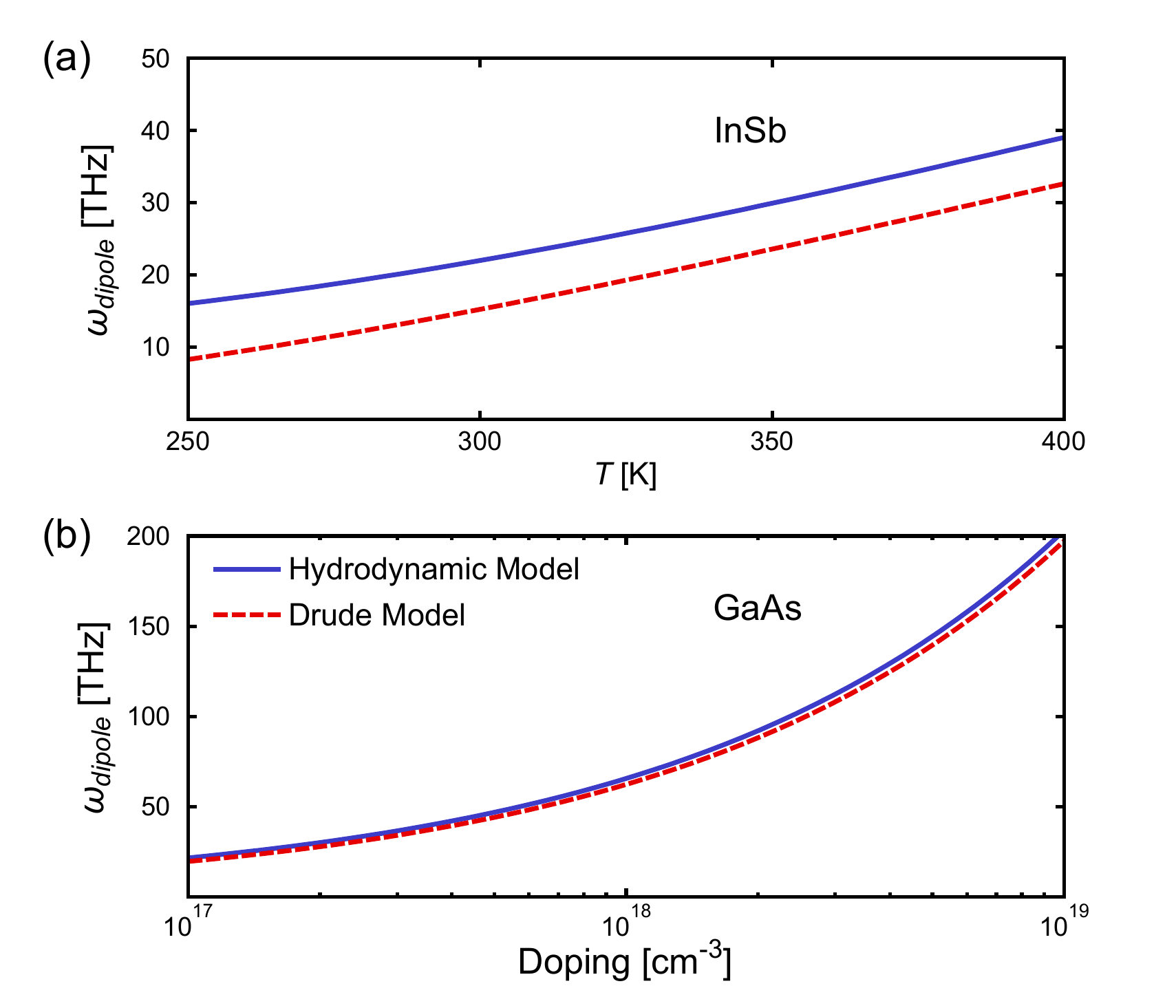}
\caption{\label{fig:temp_doping}(a) Dipole resonance frequency as a function of temperature for an InSb nanoparticle in vacuum with $R=150\mathrm{\,nm}$. (b) Dipole resonance frequency as a function of doping level in a GaAs nanoparticle in vacuum with $R=50\mathrm{\,nm}$. The dashed line is the local Drude model, and the full line is the HDM. Material parameters can be found in table \ref{tab:data}.}
\end{figure}

For particles of intrinsic InSb, the temperature will have a significant impact on the optical properties as it affects the charge carrier density and thereby the resonance frequency (as shown experimentally for a planar system in \citen{howells96}). To illustrate this, the temperature dependence of the dipole resonance in an InSb nanoparticle is shown in Fig. \ref{fig:temp_doping}(a). This time, to ensure that the results are accurate at the higher temperatures, the Fermi-Dirac distribution is used in the calculations instead of the Boltzmann distribution. As expected, the resonance frequency increases with the temperature for both the local and nonlocal solutions. This effect can be used in new plasmonic experiments where the resonance frequency is controlled within a wide range by varying the temperature. 

Such a tunability also exists in doped semiconductors where the resonance frequency instead is controlled by the doping level (as shown experimentally in \citen{garcia11}). In Fig. \ref{fig:temp_doping}(b), the dipole peak position in a GaAs nanoparticle is plotted as a function of the donor concentration, and we see how the resonance frequency goes up as the doping level increases.

The appearance of nonlocal effects in semiconductors is in a sense no surprise, as the model used is identical to the one used for metals (except the expression for $\beta$). What is really noteworthy is the magnitude of the relative blueshift. For metals, this shift is typically in the order of 5--15\% for particles of a few nm\cite{charle98,raza13b,raza15b,Scholl12}, while the blueshift seen in Fig. \ref{fig:insb_gaas}(a) is as large as 35\% despite a radius of $150\mathrm{\,nm}$. The strong blueshift is primarily explained by the small effective electron mass in InSb, which according to Eq. (\ref{eq:beta_therm}) serves to increase $\beta$. Interestingly, the relative blueshift is directly related to the non-classical fraction of the energy\cite{yan16}.

\begin{figure}
\includegraphics[width=\figwid]{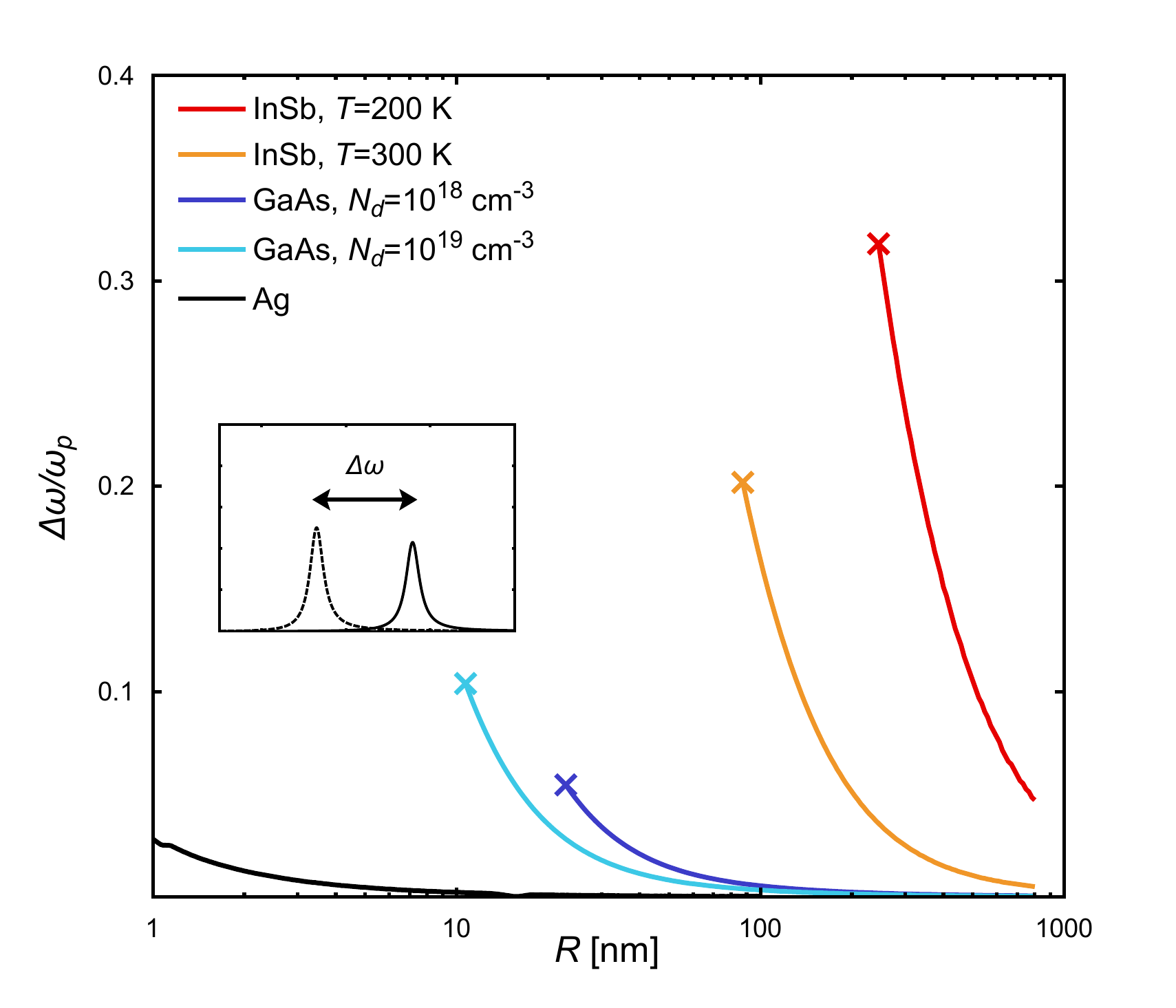}
\caption{The nonlocal blueshift $\Delta \omega$ relative to the plasma frequency $\omega_{\rm p}$, as a function of the nanosphere radius $R$. Material parameters can be found in table \ref{tab:data}. The lines are cut off with a 'x' at the left side where the particles contain fewer than 50 electrons (being a metal, silver is cut off below $1\mathrm{\,nm}$).      }
\label{fig:blueshift}
\end{figure}

To make further comparison with metal nanoparticles, the blueshift relative to the plasma frequency is in Fig.~\ref{fig:blueshift} shown  as a function of particle radius for various materials. The curves were calculated by subtracting the dipole frequency in the local model from the dipole frequency in the HDM and dividing the result by $\w_p$. The red and orange lines show the relative blueshift for InSb at $T=200\mathrm{\,K}$ and $300\mathrm{\,K}$ respectively (see the material parameters in table \ref{tab:data}). We see that the blueshift increases as the semiconductor particle becomes smaller, quite analogous to what happens for noble metals\cite{christensen14}. But unlike for metals, the curves in Fig.~\ref{fig:blueshift} also show that a lower temperature gives larger blueshifts for all semiconductor particle sizes. It has to be remembered, though, that the amplitude of the signal also decreases when the temperature is lowered, making detection harder. The 'x' at the end of each line indicates the radius where the particle contains 50 free electrons (this was the chosen lower limit of the model).

The possibility of observing nonlocal effects in semiconductors was mentioned by Hanham \etal\ in \citen{hanham12} where they studied the optical response of InSb disks with diameters of $20\mathrm{\,\upmu m}$. However, for the simulation of their results they only used the local Drude model. From Fig. \ref{fig:blueshift}, we now see that this was justified for individual InSb particles at $300\mathrm{\,K}$, as the nonlocal blueshift is negligible for radii larger than $1\mathrm{\,\upmu m}$.

The blue and pale blue lines in Fig. \ref{fig:blueshift} show the blueshifts for GaAs particles with doping levels of $10^{18}\mathrm{\,cm^{-3}}$ and $10^{19}\mathrm{\,cm^{-3}}$ respectively. Although the blueshifts are smaller than for InSb, the tendency is the same. 
 
Finally, the black line in Fig. \ref{fig:blueshift} shows the blueshift for silver particles with the parameters $\beta=1.08\times 10^{6}\mathrm{\,m/s}$, $\w_p=1.36\times 10^{16}\mathrm{\,s^{-1}}$ and $\gamma=3.80\times 10^{13}\mathrm{\,s^{-1}}$\citen{raza15}, and where $\ep_{\infty}(\omega)$ is found using the method from \citen{david11} and data from \citen{werner09}. We here see that the relative blueshift is smaller than for the semiconductors and occurs for much smaller particles. 

The hydrodynamic model is simple both conceptually and computationally, and yet it has showcased an extraordinary predictive power for the optical properties of metals. Semiconductors, however, is a new group of materials were the HDM has not yet been tested, and the situation might be more complicated. As mentioned in the Introduction, semiconductors may support excitons, an effect we have ignored here. Another phenomenon relevant for especially binary and ternary semiconductors is optical phonons which may couple to the plasmon if the resonance frequency is in the same region. This has been investigated for InSb\cite{gaur76,gu00} and GaAs\cite{olsen69,chen89}, and the mechanism could be included in the dielectric function as an extra term (as is done in \citen{jones95}).

For InSb there is yet another effect that may have to be taken into account, namely the presence of a \emph{space charge layer}. This charge carrier depleted layer stretching a few hundred angstrom into the material has been discussed in earlier papers\cite{ritz85,jones95,bell98,adomavicius09}. Such a layer would be significant for the optical properties of the InSb particle, and the question of how it would affect the nonlocal effects is still to be answered. 

The size-dependent nonlocal effects which have been investigated here would be relevant when making experimental predictions for semiconductor nanostructures in general. But semiconductors could also be used specifically for research in nonlocal effects, as the required particle sizes are much larger in semiconductors than in metals. This will be an advantage in experimental studies where the extremely small sizes of metal nanoparticles has been a challenge. Another material that also permits observation of nonlocality in larger structures than with metals is graphene. Indeed, blueshifts in arm-chair terminated graphene nanoflakes could be identified as hydrodynamic nonlocal blueshifts\cite{christensen14b}. Very recently, tunable nonlocal response of graphene has been observed in near-field imaging experiments\cite{lundberg17}.  Both graphene and semiconductors are therefore suited for research in nonlocality, as they allow the experimentalists to explore larger structures and still be able to see deviations from the local response model.

\section{Conclusions}
We have shown that size-dependent nonlocal effects are present in semiconductor particles that contain enough charge carriers to be described by the hydrodynamic Drude model. These particles are too big to behave as quantum dots, yet too small for bulk theory to apply.  Moreover, we find that the blueshift relative to the plasma frequency is much larger than what is seen in metals and that it occurs in larger particles. This finding makes semiconductors interesting and suitable candidates for further experimental explorations of nonlocal electrodynamic effects: if the required structures can be upscaled, then the fabrication is correspondingly simplified, and investigations of new, more complex geometries become realistic. 

In addition, semiconductors provide the possibility of tuning the optical response by changing the charge carrier density, for instance by temperature control and doping as investigated here. If nanoscale semiconductor structures in the future will be used in new plasmonic experiments and devices, proper modeling of the materials becomes crucial. Based on our results from the hydrodynamic model, we have clarified when nonlocality is not important and the Drude model provides sufficient description, but also when nonlocal effects should be taken into account. 

\acknowledgements
We gratefully acknowledge support from the Danish Council for Independent Research (FNU 1323-00087) and from Villum Fonden via the VKR Centre of Excellence NATEC-II. The Center for Nanostructured Graphene (CNG) was financed by the Danish National Research Council (DNRF103). N.~A.~M. is a Villum Investigator supported by Villum Fonden.

\nocite{*}
\bibliography{bib1}

\begin{thebibliography}{69}%
\makeatletter
\providecommand \@ifxundefined [1]{%
 \@ifx{#1\undefined}
}%
\providecommand \@ifnum [1]{%
 \ifnum #1\expandafter \@firstoftwo
 \else \expandafter \@secondoftwo
 \fi
}%
\providecommand \@ifx [1]{%
 \ifx #1\expandafter \@firstoftwo
 \else \expandafter \@secondoftwo
 \fi
}%
\providecommand \natexlab [1]{#1}%
\providecommand \enquote  [1]{``#1''}%
\providecommand \bibnamefont  [1]{#1}%
\providecommand \bibfnamefont [1]{#1}%
\providecommand \citenamefont [1]{#1}%
\providecommand \href@noop [0]{\@secondoftwo}%
\providecommand \href [0]{\begingroup \@sanitize@url \@href}%
\providecommand \@href[1]{\@@startlink{#1}\@@href}%
\providecommand \@@href[1]{\endgroup#1\@@endlink}%
\providecommand \@sanitize@url [0]{\catcode `\\12\catcode `\$12\catcode
  `\&12\catcode `\#12\catcode `\^12\catcode `\_12\catcode `\%12\relax}%
\providecommand \@@startlink[1]{}%
\providecommand \@@endlink[0]{}%
\providecommand \url  [0]{\begingroup\@sanitize@url \@url }%
\providecommand \@url [1]{\endgroup\@href {#1}{\urlprefix }}%
\providecommand \urlprefix  [0]{URL }%
\providecommand \Eprint [0]{\href }%
\providecommand \doibase [0]{http://dx.doi.org/}%
\providecommand \selectlanguage [0]{\@gobble}%
\providecommand \bibinfo  [0]{\@secondoftwo}%
\providecommand \bibfield  [0]{\@secondoftwo}%
\providecommand \translation [1]{[#1]}%
\providecommand \BibitemOpen [0]{}%
\providecommand \bibitemStop [0]{}%
\providecommand \bibitemNoStop [0]{.\EOS\space}%
\providecommand \EOS [0]{\spacefactor3000\relax}%
\providecommand \BibitemShut  [1]{\csname bibitem#1\endcsname}%
\let\auto@bib@innerbib\@empty
\bibitem [{\citenamefont {{Garc\'ia de Abajo}}(2008)}]{abajo08}%
  \BibitemOpen
  \bibfield  {author} {\bibinfo {author} {\bibfnamefont {F.~J.}\ \bibnamefont
  {{Garc\'ia de Abajo}}},\ }\href@noop {} {\bibfield  {journal} {\bibinfo
  {journal} {J. Phys. Chem. C}\ }\textbf {\bibinfo {volume} {112}},\ \bibinfo
  {pages} {17983} (\bibinfo {year} {2008})}\BibitemShut {NoStop}%
\bibitem [{\citenamefont {McMahon}\ \emph {et~al.}(2009)\citenamefont
  {McMahon}, \citenamefont {Gray},\ and\ \citenamefont {Schatz}}]{mcmahon09}%
  \BibitemOpen
  \bibfield  {author} {\bibinfo {author} {\bibfnamefont {J.~M.}\ \bibnamefont
  {McMahon}}, \bibinfo {author} {\bibfnamefont {S.~K.}\ \bibnamefont {Gray}}, \
  and\ \bibinfo {author} {\bibfnamefont {G.~C.}\ \bibnamefont {Schatz}},\
  }\href@noop {} {\bibfield  {journal} {\bibinfo  {journal} {Phys. Rev. Lett.}\
  }\textbf {\bibinfo {volume} {103}},\ \bibinfo {pages} {097403} (\bibinfo
  {year} {2009})}\BibitemShut {NoStop}%
\bibitem [{\citenamefont {Raza}\ \emph
  {et~al.}(2013{\natexlab{a}})\citenamefont {Raza}, \citenamefont {Yan},
  \citenamefont {Stenger}, \citenamefont {Wubs},\ and\ \citenamefont
  {Mortensen}}]{raza13}%
  \BibitemOpen
  \bibfield  {author} {\bibinfo {author} {\bibfnamefont {S.}~\bibnamefont
  {Raza}}, \bibinfo {author} {\bibfnamefont {W.}~\bibnamefont {Yan}}, \bibinfo
  {author} {\bibfnamefont {N.}~\bibnamefont {Stenger}}, \bibinfo {author}
  {\bibfnamefont {M.}~\bibnamefont {Wubs}}, \ and\ \bibinfo {author}
  {\bibfnamefont {N.~A.}\ \bibnamefont {Mortensen}},\ }\href@noop {} {\bibfield
   {journal} {\bibinfo  {journal} {Opt. Express}\ }\textbf {\bibinfo {volume}
  {21}},\ \bibinfo {pages} {27344} (\bibinfo {year}
  {2013}{\natexlab{a}})}\BibitemShut {NoStop}%
\bibitem [{\citenamefont {Toscano}\ \emph {et~al.}(2015)\citenamefont
  {Toscano}, \citenamefont {Straubel}, \citenamefont {Kwiatkowski},
  \citenamefont {Rockstuhl}, \citenamefont {Evers}, \citenamefont {Xu},
  \citenamefont {Mortensen},\ and\ \citenamefont {Wubs}}]{toscano15}%
  \BibitemOpen
  \bibfield  {author} {\bibinfo {author} {\bibfnamefont {G.}~\bibnamefont
  {Toscano}}, \bibinfo {author} {\bibfnamefont {J.}~\bibnamefont {Straubel}},
  \bibinfo {author} {\bibfnamefont {A.}~\bibnamefont {Kwiatkowski}}, \bibinfo
  {author} {\bibfnamefont {C.}~\bibnamefont {Rockstuhl}}, \bibinfo {author}
  {\bibfnamefont {F.}~\bibnamefont {Evers}}, \bibinfo {author} {\bibfnamefont
  {H.}~\bibnamefont {Xu}}, \bibinfo {author} {\bibfnamefont {N.~A.}\
  \bibnamefont {Mortensen}}, \ and\ \bibinfo {author} {\bibfnamefont
  {M.}~\bibnamefont {Wubs}},\ }\href@noop {} {\bibfield  {journal} {\bibinfo
  {journal} {Nat. Commun.}\ }\textbf {\bibinfo {volume} {6}},\ \bibinfo {pages}
  {1} (\bibinfo {year} {2015})}\BibitemShut {NoStop}%
\bibitem [{\citenamefont {Christensen}\ \emph
  {et~al.}(2014{\natexlab{a}})\citenamefont {Christensen}, \citenamefont {Yan},
  \citenamefont {Raza}, \citenamefont {Jauho}, \citenamefont {Mortsensen},\
  and\ \citenamefont {Wubs}}]{christensen14}%
  \BibitemOpen
  \bibfield  {author} {\bibinfo {author} {\bibfnamefont {T.}~\bibnamefont
  {Christensen}}, \bibinfo {author} {\bibfnamefont {W.}~\bibnamefont {Yan}},
  \bibinfo {author} {\bibfnamefont {S.}~\bibnamefont {Raza}}, \bibinfo {author}
  {\bibfnamefont {A.-P.}\ \bibnamefont {Jauho}}, \bibinfo {author}
  {\bibfnamefont {N.~A.}\ \bibnamefont {Mortsensen}}, \ and\ \bibinfo {author}
  {\bibfnamefont {M.}~\bibnamefont {Wubs}},\ }\href@noop {} {\bibfield
  {journal} {\bibinfo  {journal} {ACS Nano}\ }\textbf {\bibinfo {volume} {8}},\
  \bibinfo {pages} {1745} (\bibinfo {year} {2014}{\natexlab{a}})}\BibitemShut
  {NoStop}%
\bibitem [{\citenamefont {Raza}\ \emph
  {et~al.}(2015{\natexlab{a}})\citenamefont {Raza}, \citenamefont
  {Bozhevolnyi}, \citenamefont {Wubs},\ and\ \citenamefont
  {Mortensen}}]{raza15}%
  \BibitemOpen
  \bibfield  {author} {\bibinfo {author} {\bibfnamefont {S.}~\bibnamefont
  {Raza}}, \bibinfo {author} {\bibfnamefont {S.~I.}\ \bibnamefont
  {Bozhevolnyi}}, \bibinfo {author} {\bibfnamefont {M.}~\bibnamefont {Wubs}}, \
  and\ \bibinfo {author} {\bibfnamefont {N.~A.}\ \bibnamefont {Mortensen}},\
  }\href@noop {} {\bibfield  {journal} {\bibinfo  {journal} {J. Phys.: Condens.
  Mat.}\ }\textbf {\bibinfo {volume} {27}},\ \bibinfo {pages} {183204}
  (\bibinfo {year} {2015}{\natexlab{a}})}\BibitemShut {NoStop}%
\bibitem [{\citenamefont {Fox}(2010)}]{Fox}%
  \BibitemOpen
  \bibfield  {author} {\bibinfo {author} {\bibfnamefont {M.}~\bibnamefont
  {Fox}},\ }\href@noop {} {\emph {\bibinfo {title} {Optical Properties of
  Solids}}},\ \bibinfo {edition} {2nd}\ ed.\ (\bibinfo  {publisher} {Oxford
  University Press},\ \bibinfo {year} {2010})\BibitemShut {NoStop}%
\bibitem [{\citenamefont {Hanham}\ \emph {et~al.}(2012)\citenamefont {Hanham},
  \citenamefont {Fernández-Domínguez}, \citenamefont {Teng}, \citenamefont
  {Ang}, \citenamefont {Lim}, \citenamefont {Yoon}, \citenamefont {Ngo},
  \citenamefont {Klein}, \citenamefont {Pendry},\ and\ \citenamefont
  {Maier}}]{hanham12}%
  \BibitemOpen
  \bibfield  {author} {\bibinfo {author} {\bibfnamefont {S.~M.}\ \bibnamefont
  {Hanham}}, \bibinfo {author} {\bibfnamefont {A.~I.}\ \bibnamefont
  {Fernández-Domínguez}}, \bibinfo {author} {\bibfnamefont {J.~H.}\
  \bibnamefont {Teng}}, \bibinfo {author} {\bibfnamefont {S.~S.}\ \bibnamefont
  {Ang}}, \bibinfo {author} {\bibfnamefont {K.~P.}\ \bibnamefont {Lim}},
  \bibinfo {author} {\bibfnamefont {S.~F.}\ \bibnamefont {Yoon}}, \bibinfo
  {author} {\bibfnamefont {C.~Y.}\ \bibnamefont {Ngo}}, \bibinfo {author}
  {\bibfnamefont {N.}~\bibnamefont {Klein}}, \bibinfo {author} {\bibfnamefont
  {J.~B.}\ \bibnamefont {Pendry}}, \ and\ \bibinfo {author} {\bibfnamefont
  {S.~A.}\ \bibnamefont {Maier}},\ }\href@noop {} {\bibfield  {journal}
  {\bibinfo  {journal} {Adv. Mater.}\ }\textbf {\bibinfo {volume} {24}},\
  \bibinfo {pages} {226} (\bibinfo {year} {2012})}\BibitemShut {NoStop}%
\bibitem [{\citenamefont {Haug}\ and\ \citenamefont {Koch}(2009)}]{haug}%
  \BibitemOpen
  \bibfield  {author} {\bibinfo {author} {\bibfnamefont {H.}~\bibnamefont
  {Haug}}\ and\ \bibinfo {author} {\bibfnamefont {S.~W.}\ \bibnamefont
  {Koch}},\ }\href@noop {} {\emph {\bibinfo {title} {Quantum theory of the
  optical and electronic properties of semiconductors}}},\ \bibinfo {edition}
  {5th}\ ed.\ (\bibinfo  {publisher} {World Scientific},\ \bibinfo {year}
  {2009})\BibitemShut {NoStop}%
\bibitem [{\citenamefont {Kelly}(1995)}]{kelly}%
  \BibitemOpen
  \bibfield  {author} {\bibinfo {author} {\bibfnamefont {M.~J.}\ \bibnamefont
  {Kelly}},\ }\href@noop {} {\emph {\bibinfo {title} {Low-dimensional
  semiconductors}}},\ \bibinfo {edition} {1st}\ ed.\ (\bibinfo  {publisher}
  {Clarendon Press},\ \bibinfo {year} {1995})\BibitemShut {NoStop}%
\bibitem [{\citenamefont {Ritz}\ and\ \citenamefont {Lüth}(1984)}]{ritz84}%
  \BibitemOpen
  \bibfield  {author} {\bibinfo {author} {\bibfnamefont {A.}~\bibnamefont
  {Ritz}}\ and\ \bibinfo {author} {\bibfnamefont {H.}~\bibnamefont {Lüth}},\
  }\href@noop {} {\bibfield  {journal} {\bibinfo  {journal} {Phys. Rev. Lett.}\
  }\textbf {\bibinfo {volume} {52}},\ \bibinfo {pages} {1242} (\bibinfo {year}
  {1984})}\BibitemShut {NoStop}%
\bibitem [{\citenamefont {Jones}\ \emph {et~al.}(1995)\citenamefont {Jones},
  \citenamefont {Schweitzer}, \citenamefont {Richardson}, \citenamefont
  {Bell},\ and\ \citenamefont {McConville}}]{jones95}%
  \BibitemOpen
  \bibfield  {author} {\bibinfo {author} {\bibfnamefont {T.~S.}\ \bibnamefont
  {Jones}}, \bibinfo {author} {\bibfnamefont {M.~O.}\ \bibnamefont
  {Schweitzer}}, \bibinfo {author} {\bibfnamefont {N.~V.}\ \bibnamefont
  {Richardson}}, \bibinfo {author} {\bibfnamefont {G.~R.}\ \bibnamefont
  {Bell}}, \ and\ \bibinfo {author} {\bibfnamefont {C.~F.}\ \bibnamefont
  {McConville}},\ }\href@noop {} {\bibfield  {journal} {\bibinfo  {journal}
  {Phys. Rev. B}\ }\textbf {\bibinfo {volume} {51}},\ \bibinfo {pages} {17675}
  (\bibinfo {year} {1995})}\BibitemShut {NoStop}%
\bibitem [{\citenamefont {Rivas}\ \emph {et~al.}(1996)\citenamefont {Rivas},
  \citenamefont {Kuttge}, \citenamefont {Kurz}, \citenamefont {Bolivar},\ and\
  \citenamefont {Sánchez-Gil}}]{rivas06}%
  \BibitemOpen
  \bibfield  {author} {\bibinfo {author} {\bibfnamefont {J.~G.}\ \bibnamefont
  {Rivas}}, \bibinfo {author} {\bibfnamefont {M.}~\bibnamefont {Kuttge}},
  \bibinfo {author} {\bibfnamefont {H.}~\bibnamefont {Kurz}}, \bibinfo {author}
  {\bibfnamefont {P.~H.}\ \bibnamefont {Bolivar}}, \ and\ \bibinfo {author}
  {\bibfnamefont {J.~A.}\ \bibnamefont {Sánchez-Gil}},\ }\href@noop {}
  {\bibfield  {journal} {\bibinfo  {journal} {Appl. Phys. Lett.}\ }\textbf
  {\bibinfo {volume} {88}},\ \bibinfo {pages} {082106} (\bibinfo {year}
  {1996})}\BibitemShut {NoStop}%
\bibitem [{\citenamefont {Naik}\ \emph {et~al.}(2013)\citenamefont {Naik},
  \citenamefont {Shalaev},\ and\ \citenamefont {Boltasseva}}]{naik13}%
  \BibitemOpen
  \bibfield  {author} {\bibinfo {author} {\bibfnamefont {G.~V.}\ \bibnamefont
  {Naik}}, \bibinfo {author} {\bibfnamefont {V.~M.}\ \bibnamefont {Shalaev}}, \
  and\ \bibinfo {author} {\bibfnamefont {A.}~\bibnamefont {Boltasseva}},\
  }\href@noop {} {\bibfield  {journal} {\bibinfo  {journal} {Adv. Mater.}\
  }\textbf {\bibinfo {volume} {25}},\ \bibinfo {pages} {3264} (\bibinfo {year}
  {2013})}\BibitemShut {NoStop}%
\bibitem [{\citenamefont {Matz}\ and\ \citenamefont {Lüth}(1981)}]{matz81}%
  \BibitemOpen
  \bibfield  {author} {\bibinfo {author} {\bibfnamefont {R.}~\bibnamefont
  {Matz}}\ and\ \bibinfo {author} {\bibfnamefont {H.}~\bibnamefont {Lüth}},\
  }\href@noop {} {\bibfield  {journal} {\bibinfo  {journal} {Phys. Rev. Lett.}\
  }\textbf {\bibinfo {volume} {46}},\ \bibinfo {pages} {500} (\bibinfo {year}
  {1981})}\BibitemShut {NoStop}%
\bibitem [{\citenamefont {Gray-Grychowski}\ \emph {et~al.}(1986)\citenamefont
  {Gray-Grychowski}, \citenamefont {Stradling}, \citenamefont {Egdell},
  \citenamefont {Dobson}, \citenamefont {Joyce},\ and\ \citenamefont
  {Woodbridge}}]{grychowski86}%
  \BibitemOpen
  \bibfield  {author} {\bibinfo {author} {\bibfnamefont {Z.~J.}\ \bibnamefont
  {Gray-Grychowski}}, \bibinfo {author} {\bibfnamefont {R.~A.}\ \bibnamefont
  {Stradling}}, \bibinfo {author} {\bibfnamefont {R.~G.}\ \bibnamefont
  {Egdell}}, \bibinfo {author} {\bibfnamefont {P.~J.}\ \bibnamefont {Dobson}},
  \bibinfo {author} {\bibfnamefont {B.~A.}\ \bibnamefont {Joyce}}, \ and\
  \bibinfo {author} {\bibfnamefont {K.}~\bibnamefont {Woodbridge}},\
  }\href@noop {} {\bibfield  {journal} {\bibinfo  {journal} {solid state
  comm.}\ }\textbf {\bibinfo {volume} {59}},\ \bibinfo {pages} {703} (\bibinfo
  {year} {1986})}\BibitemShut {NoStop}%
\bibitem [{\citenamefont {Betti}\ \emph {et~al.}(1989)\citenamefont {Betti},
  \citenamefont {del Pennino},\ and\ \citenamefont {Mariani}}]{betti89}%
  \BibitemOpen
  \bibfield  {author} {\bibinfo {author} {\bibfnamefont {M.~G.}\ \bibnamefont
  {Betti}}, \bibinfo {author} {\bibfnamefont {U.}~\bibnamefont {del Pennino}},
  \ and\ \bibinfo {author} {\bibfnamefont {C.}~\bibnamefont {Mariani}},\
  }\href@noop {} {\bibfield  {journal} {\bibinfo  {journal} {Phys. Rev. B}\
  }\textbf {\bibinfo {volume} {39}},\ \bibinfo {pages} {5887} (\bibinfo {year}
  {1989})}\BibitemShut {NoStop}%
\bibitem [{\citenamefont {Chen}\ \emph {et~al.}(1989)\citenamefont {Chen},
  \citenamefont {Nannarone}, \citenamefont {Schaefer}, \citenamefont
  {Hermanson},\ and\ \citenamefont {Lapeyre}}]{chen89}%
  \BibitemOpen
  \bibfield  {author} {\bibinfo {author} {\bibfnamefont {Y.}~\bibnamefont
  {Chen}}, \bibinfo {author} {\bibfnamefont {S.}~\bibnamefont {Nannarone}},
  \bibinfo {author} {\bibfnamefont {J.}~\bibnamefont {Schaefer}}, \bibinfo
  {author} {\bibfnamefont {J.~C.}\ \bibnamefont {Hermanson}}, \ and\ \bibinfo
  {author} {\bibfnamefont {G.~J.}\ \bibnamefont {Lapeyre}},\ }\href@noop {}
  {\bibfield  {journal} {\bibinfo  {journal} {Phys. Rev. B}\ }\textbf {\bibinfo
  {volume} {39}},\ \bibinfo {pages} {7653} (\bibinfo {year}
  {1989})}\BibitemShut {NoStop}%
\bibitem [{\citenamefont {Meng}\ \emph {et~al.}(1991)\citenamefont {Meng},
  \citenamefont {Anderson}, \citenamefont {Hermanson},\ and\ \citenamefont
  {Lapeyre}}]{meng91}%
  \BibitemOpen
  \bibfield  {author} {\bibinfo {author} {\bibfnamefont {Y.}~\bibnamefont
  {Meng}}, \bibinfo {author} {\bibfnamefont {J.~R.}\ \bibnamefont {Anderson}},
  \bibinfo {author} {\bibfnamefont {J.~C.}\ \bibnamefont {Hermanson}}, \ and\
  \bibinfo {author} {\bibfnamefont {G.~J.}\ \bibnamefont {Lapeyre}},\
  }\href@noop {} {\bibfield  {journal} {\bibinfo  {journal} {Phys. Rev. B}\
  }\textbf {\bibinfo {volume} {44}},\ \bibinfo {pages} {4040} (\bibinfo {year}
  {1991})}\BibitemShut {NoStop}%
\bibitem [{\citenamefont {Biagi}\ \emph {et~al.}(1992)\citenamefont {Biagi},
  \citenamefont {Mariani},\ and\ \citenamefont {del Pennino}}]{biagi92}%
  \BibitemOpen
  \bibfield  {author} {\bibinfo {author} {\bibfnamefont {R.}~\bibnamefont
  {Biagi}}, \bibinfo {author} {\bibfnamefont {C.}~\bibnamefont {Mariani}}, \
  and\ \bibinfo {author} {\bibfnamefont {U.}~\bibnamefont {del Pennino}},\
  }\href@noop {} {\bibfield  {journal} {\bibinfo  {journal} {Phys. Rev. B}\
  }\textbf {\bibinfo {volume} {46}},\ \bibinfo {pages} {2467} (\bibinfo {year}
  {1992})}\BibitemShut {NoStop}%
\bibitem [{\citenamefont {Ginn}\ \emph {et~al.}(2011)\citenamefont {Ginn},
  \citenamefont {{Jarecki Jr.}}, \citenamefont {Shaner},\ and\ \citenamefont
  {Davids}}]{ginn11}%
  \BibitemOpen
  \bibfield  {author} {\bibinfo {author} {\bibfnamefont {J.~C.}\ \bibnamefont
  {Ginn}}, \bibinfo {author} {\bibfnamefont {R.~L.}\ \bibnamefont {{Jarecki
  Jr.}}}, \bibinfo {author} {\bibfnamefont {E.~A.}\ \bibnamefont {Shaner}}, \
  and\ \bibinfo {author} {\bibfnamefont {P.~S.}\ \bibnamefont {Davids}},\
  }\href@noop {} {\bibfield  {journal} {\bibinfo  {journal} {J. Appl. Phys.}\
  }\textbf {\bibinfo {volume} {110}},\ \bibinfo {pages} {043110} (\bibinfo
  {year} {2011})}\BibitemShut {NoStop}%
\bibitem [{\citenamefont {Luther}\ \emph {et~al.}(2011)\citenamefont {Luther},
  \citenamefont {Jain}, \citenamefont {Ewers},\ and\ \citenamefont
  {Alivisatos}}]{luther11}%
  \BibitemOpen
  \bibfield  {author} {\bibinfo {author} {\bibfnamefont {J.~M.}\ \bibnamefont
  {Luther}}, \bibinfo {author} {\bibfnamefont {P.~K.}\ \bibnamefont {Jain}},
  \bibinfo {author} {\bibfnamefont {T.}~\bibnamefont {Ewers}}, \ and\ \bibinfo
  {author} {\bibfnamefont {A.~P.}\ \bibnamefont {Alivisatos}},\ }\href@noop {}
  {\bibfield  {journal} {\bibinfo  {journal} {Nat. Mat.}\ }\textbf {\bibinfo
  {volume} {10}},\ \bibinfo {pages} {361} (\bibinfo {year} {2011})}\BibitemShut
  {NoStop}%
\bibitem [{\citenamefont {Garcia}\ \emph {et~al.}(2011)\citenamefont {Garcia},
  \citenamefont {Buonsanti}, \citenamefont {Runnerstrom}, \citenamefont
  {Mendelsberg}, \citenamefont {Llordes}, \citenamefont {Anders}, \citenamefont
  {Richardson},\ and\ \citenamefont {Milliron}}]{garcia11}%
  \BibitemOpen
  \bibfield  {author} {\bibinfo {author} {\bibfnamefont {G.}~\bibnamefont
  {Garcia}}, \bibinfo {author} {\bibfnamefont {R.}~\bibnamefont {Buonsanti}},
  \bibinfo {author} {\bibfnamefont {E.~L.}\ \bibnamefont {Runnerstrom}},
  \bibinfo {author} {\bibfnamefont {R.~J.}\ \bibnamefont {Mendelsberg}},
  \bibinfo {author} {\bibfnamefont {A.}~\bibnamefont {Llordes}}, \bibinfo
  {author} {\bibfnamefont {A.}~\bibnamefont {Anders}}, \bibinfo {author}
  {\bibfnamefont {T.~J.}\ \bibnamefont {Richardson}}, \ and\ \bibinfo {author}
  {\bibfnamefont {D.~J.}\ \bibnamefont {Milliron}},\ }\href@noop {} {\bibfield
  {journal} {\bibinfo  {journal} {Nano Lett.}\ }\textbf {\bibinfo {volume}
  {11}},\ \bibinfo {pages} {4415} (\bibinfo {year} {2011})}\BibitemShut
  {NoStop}%
\bibitem [{\citenamefont {Buonsanti}\ \emph {et~al.}(2011)\citenamefont
  {Buonsanti}, \citenamefont {Llordes}, \citenamefont {Aloni}, \citenamefont
  {Helms},\ and\ \citenamefont {Milliron}}]{buonsanti11}%
  \BibitemOpen
  \bibfield  {author} {\bibinfo {author} {\bibfnamefont {R.}~\bibnamefont
  {Buonsanti}}, \bibinfo {author} {\bibfnamefont {A.}~\bibnamefont {Llordes}},
  \bibinfo {author} {\bibfnamefont {S.}~\bibnamefont {Aloni}}, \bibinfo
  {author} {\bibfnamefont {B.~A.}\ \bibnamefont {Helms}}, \ and\ \bibinfo
  {author} {\bibfnamefont {D.~J.}\ \bibnamefont {Milliron}},\ }\href@noop {}
  {\bibfield  {journal} {\bibinfo  {journal} {Nano Lett.}\ }\textbf {\bibinfo
  {volume} {11}},\ \bibinfo {pages} {4706} (\bibinfo {year}
  {2011})}\BibitemShut {NoStop}%
\bibitem [{\citenamefont {Manthiram}\ and\ \citenamefont
  {Alivisatos}(2012)}]{manthiram12}%
  \BibitemOpen
  \bibfield  {author} {\bibinfo {author} {\bibfnamefont {K.}~\bibnamefont
  {Manthiram}}\ and\ \bibinfo {author} {\bibfnamefont {A.~P.}\ \bibnamefont
  {Alivisatos}},\ }\href@noop {} {\bibfield  {journal} {\bibinfo  {journal} {J.
  Am. Chem. Soc.}\ }\textbf {\bibinfo {volume} {134}},\ \bibinfo {pages} {3995}
  (\bibinfo {year} {2012})}\BibitemShut {NoStop}%
\bibitem [{\citenamefont {Schimpf}\ \emph {et~al.}(2014)\citenamefont
  {Schimpf}, \citenamefont {Thakkar}, \citenamefont {Gunthardt}, \citenamefont
  {Masiello},\ and\ \citenamefont {Gamelin}}]{schimpf13}%
  \BibitemOpen
  \bibfield  {author} {\bibinfo {author} {\bibfnamefont {A.~M.}\ \bibnamefont
  {Schimpf}}, \bibinfo {author} {\bibfnamefont {N.}~\bibnamefont {Thakkar}},
  \bibinfo {author} {\bibfnamefont {C.~E.}\ \bibnamefont {Gunthardt}}, \bibinfo
  {author} {\bibfnamefont {D.~J.}\ \bibnamefont {Masiello}}, \ and\ \bibinfo
  {author} {\bibfnamefont {D.~R.}\ \bibnamefont {Gamelin}},\ }\href@noop {}
  {\bibfield  {journal} {\bibinfo  {journal} {ACS Nano}\ }\textbf {\bibinfo
  {volume} {8}},\ \bibinfo {pages} {1065} (\bibinfo {year} {2014})}\BibitemShut
  {NoStop}%
\bibitem [{\citenamefont {Zhou}\ \emph {et~al.}(2015)\citenamefont {Zhou},
  \citenamefont {Pi}, \citenamefont {Ni}, \citenamefont {Ding}, \citenamefont
  {Jiang}, \citenamefont {Jin}, \citenamefont {Delerue}, \citenamefont {Yang},\
  and\ \citenamefont {Nozaki}}]{zhou15}%
  \BibitemOpen
  \bibfield  {author} {\bibinfo {author} {\bibfnamefont {S.}~\bibnamefont
  {Zhou}}, \bibinfo {author} {\bibfnamefont {X.}~\bibnamefont {Pi}}, \bibinfo
  {author} {\bibfnamefont {Z.}~\bibnamefont {Ni}}, \bibinfo {author}
  {\bibfnamefont {Y.}~\bibnamefont {Ding}}, \bibinfo {author} {\bibfnamefont
  {Y.}~\bibnamefont {Jiang}}, \bibinfo {author} {\bibfnamefont
  {C.}~\bibnamefont {Jin}}, \bibinfo {author} {\bibfnamefont {C.}~\bibnamefont
  {Delerue}}, \bibinfo {author} {\bibfnamefont {D.}~\bibnamefont {Yang}}, \
  and\ \bibinfo {author} {\bibfnamefont {T.}~\bibnamefont {Nozaki}},\
  }\href@noop {} {\bibfield  {journal} {\bibinfo  {journal} {ACS Nano}\
  }\textbf {\bibinfo {volume} {9}},\ \bibinfo {pages} {378} (\bibinfo {year}
  {2015})}\BibitemShut {NoStop}%
\bibitem [{\citenamefont {Hapala}\ \emph {et~al.}(2013)\citenamefont {Hapala},
  \citenamefont {Kusov\'a}, \citenamefont {Pelant},\ and\ \citenamefont
  {Jel\'inek}}]{hapala13}%
  \BibitemOpen
  \bibfield  {author} {\bibinfo {author} {\bibfnamefont {P.}~\bibnamefont
  {Hapala}}, \bibinfo {author} {\bibfnamefont {K.}~\bibnamefont {Kusov\'a}},
  \bibinfo {author} {\bibfnamefont {I.}~\bibnamefont {Pelant}}, \ and\ \bibinfo
  {author} {\bibfnamefont {P.}~\bibnamefont {Jel\'inek}},\ }\href@noop {}
  {\bibfield  {journal} {\bibinfo  {journal} {Phys. Rev. B}\ }\textbf {\bibinfo
  {volume} {87}},\ \bibinfo {pages} {195420} (\bibinfo {year}
  {2013})}\BibitemShut {NoStop}%
\bibitem [{\citenamefont {Zhang}\ \emph {et~al.}(2014)\citenamefont {Zhang},
  \citenamefont {Kulkarni}, \citenamefont {Prodan}, \citenamefont
  {Nordlander},\ and\ \citenamefont {Govorov}}]{zhang14}%
  \BibitemOpen
  \bibfield  {author} {\bibinfo {author} {\bibfnamefont {H.}~\bibnamefont
  {Zhang}}, \bibinfo {author} {\bibfnamefont {V.}~\bibnamefont {Kulkarni}},
  \bibinfo {author} {\bibfnamefont {E.}~\bibnamefont {Prodan}}, \bibinfo
  {author} {\bibfnamefont {P.}~\bibnamefont {Nordlander}}, \ and\ \bibinfo
  {author} {\bibfnamefont {A.~O.}\ \bibnamefont {Govorov}},\ }\href@noop {}
  {\bibfield  {journal} {\bibinfo  {journal} {J. Phys. Chem. C}\ }\textbf
  {\bibinfo {volume} {118}},\ \bibinfo {pages} {16035} (\bibinfo {year}
  {2014})}\BibitemShut {NoStop}%
\bibitem [{\citenamefont {Faucheaux}\ \emph {et~al.}(2014)\citenamefont
  {Faucheaux}, \citenamefont {Stanton},\ and\ \citenamefont
  {Jain}}]{faucheaux14}%
  \BibitemOpen
  \bibfield  {author} {\bibinfo {author} {\bibfnamefont {J.~A.}\ \bibnamefont
  {Faucheaux}}, \bibinfo {author} {\bibfnamefont {A.~L.~D.}\ \bibnamefont
  {Stanton}}, \ and\ \bibinfo {author} {\bibfnamefont {P.~K.}\ \bibnamefont
  {Jain}},\ }\href@noop {} {\bibfield  {journal} {\bibinfo  {journal} {J. Phys.
  Chem. Lett.}\ }\textbf {\bibinfo {volume} {5}},\ \bibinfo {pages} {976}
  (\bibinfo {year} {2014})}\BibitemShut {NoStop}%
\bibitem [{\citenamefont {Monreal}\ \emph {et~al.}(2015)\citenamefont
  {Monreal}, \citenamefont {Antosiewicz},\ and\ \citenamefont
  {Apell}}]{carmina15}%
  \BibitemOpen
  \bibfield  {author} {\bibinfo {author} {\bibfnamefont {R.~C.}\ \bibnamefont
  {Monreal}}, \bibinfo {author} {\bibfnamefont {T.~J.}\ \bibnamefont
  {Antosiewicz}}, \ and\ \bibinfo {author} {\bibfnamefont {S.~P.}\ \bibnamefont
  {Apell}},\ }\href@noop {} {\bibfield  {journal} {\bibinfo  {journal} {J.
  Phys. Chem. Lett.}\ }\textbf {\bibinfo {volume} {6}},\ \bibinfo {pages}
  {1847} (\bibinfo {year} {2015})}\BibitemShut {NoStop}%
\bibitem [{\citenamefont {Zhang}\ \emph {et~al.}(2017)\citenamefont {Zhang},
  \citenamefont {Zhang}, \citenamefont {Schramke}, \citenamefont {Bedford},
  \citenamefont {Hunter}, \citenamefont {Kortshagen},\ and\ \citenamefont
  {Nordlander}}]{zhang17}%
  \BibitemOpen
  \bibfield  {author} {\bibinfo {author} {\bibfnamefont {H.}~\bibnamefont
  {Zhang}}, \bibinfo {author} {\bibfnamefont {R.}~\bibnamefont {Zhang}},
  \bibinfo {author} {\bibfnamefont {K.~S.}\ \bibnamefont {Schramke}}, \bibinfo
  {author} {\bibfnamefont {N.~M.}\ \bibnamefont {Bedford}}, \bibinfo {author}
  {\bibfnamefont {K.}~\bibnamefont {Hunter}}, \bibinfo {author} {\bibfnamefont
  {U.~R.}\ \bibnamefont {Kortshagen}}, \ and\ \bibinfo {author} {\bibfnamefont
  {P.}~\bibnamefont {Nordlander}},\ }\href {\doibase
  10.1021/acsphotonics.7b00026} {\bibfield  {journal} {\bibinfo  {journal} {ACS
  Photonics}\ }\textbf {\bibinfo {volume} {4}},\ \bibinfo {pages} {963}
  (\bibinfo {year} {2017})}\BibitemShut {NoStop}%
\bibitem [{\citenamefont {Wubs}(2015)}]{wubs15}%
  \BibitemOpen
  \bibfield  {author} {\bibinfo {author} {\bibfnamefont {M.}~\bibnamefont
  {Wubs}},\ }\href@noop {} {\bibfield  {journal} {\bibinfo  {journal} {Opt.
  Expr.}\ }\textbf {\bibinfo {volume} {23}},\ \bibinfo {pages} {31296}
  (\bibinfo {year} {2015})}\BibitemShut {NoStop}%
\bibitem [{\citenamefont {Raza}\ \emph
  {et~al.}(2013{\natexlab{b}})\citenamefont {Raza}, \citenamefont
  {Christensen}, \citenamefont {Wubs}, \citenamefont {Bozhevolnyi},\ and\
  \citenamefont {Mortensen}}]{Raza13a}%
  \BibitemOpen
  \bibfield  {author} {\bibinfo {author} {\bibfnamefont {S.}~\bibnamefont
  {Raza}}, \bibinfo {author} {\bibfnamefont {T.}~\bibnamefont {Christensen}},
  \bibinfo {author} {\bibfnamefont {M.}~\bibnamefont {Wubs}}, \bibinfo {author}
  {\bibfnamefont {S.~I.}\ \bibnamefont {Bozhevolnyi}}, \ and\ \bibinfo {author}
  {\bibfnamefont {N.~A.}\ \bibnamefont {Mortensen}},\ }\href@noop {} {\bibfield
   {journal} {\bibinfo  {journal} {Phys. Rev. B}\ }\textbf {\bibinfo {volume}
  {88}},\ \bibinfo {pages} {115401} (\bibinfo {year}
  {2013}{\natexlab{b}})}\BibitemShut {NoStop}%
\bibitem [{\citenamefont {Sze}(1969)}]{sze}%
  \BibitemOpen
  \bibfield  {author} {\bibinfo {author} {\bibfnamefont {S.~M.}\ \bibnamefont
  {Sze}},\ }\href@noop {} {\emph {\bibinfo {title} {Physics of semiconductor
  devices}}},\ \bibinfo {edition} {1st}\ ed.\ (\bibinfo  {publisher} {Wiley},\
  \bibinfo {year} {1969})\BibitemShut {NoStop}%
\bibitem [{\citenamefont {Raza}\ \emph {et~al.}(2011)\citenamefont {Raza},
  \citenamefont {Toscano}, \citenamefont {Jauho}, \citenamefont {Wubs},\ and\
  \citenamefont {Mortensen}}]{raza11}%
  \BibitemOpen
  \bibfield  {author} {\bibinfo {author} {\bibfnamefont {S.}~\bibnamefont
  {Raza}}, \bibinfo {author} {\bibfnamefont {G.}~\bibnamefont {Toscano}},
  \bibinfo {author} {\bibfnamefont {A.-P.}\ \bibnamefont {Jauho}}, \bibinfo
  {author} {\bibfnamefont {M.}~\bibnamefont {Wubs}}, \ and\ \bibinfo {author}
  {\bibfnamefont {N.~A.}\ \bibnamefont {Mortensen}},\ }\href {\doibase
  10.1103/PhysRevB.84.121412} {\bibfield  {journal} {\bibinfo  {journal} {Phys.
  Rev. B}\ }\textbf {\bibinfo {volume} {84}},\ \bibinfo {pages} {121412}
  (\bibinfo {year} {2011})}\BibitemShut {NoStop}%
\bibitem [{\citenamefont {Toscano}\ \emph {et~al.}(2012)\citenamefont
  {Toscano}, \citenamefont {Raza}, \citenamefont {Jauho}, \citenamefont
  {Mortensen},\ and\ \citenamefont {Wubs}}]{toscano12}%
  \BibitemOpen
  \bibfield  {author} {\bibinfo {author} {\bibfnamefont {G.}~\bibnamefont
  {Toscano}}, \bibinfo {author} {\bibfnamefont {S.}~\bibnamefont {Raza}},
  \bibinfo {author} {\bibfnamefont {A.-P.}\ \bibnamefont {Jauho}}, \bibinfo
  {author} {\bibfnamefont {N.~A.}\ \bibnamefont {Mortensen}}, \ and\ \bibinfo
  {author} {\bibfnamefont {M.}~\bibnamefont {Wubs}},\ }\href {\doibase
  10.1364/OE.20.004176} {\bibfield  {journal} {\bibinfo  {journal} {Opt.
  Express}\ }\textbf {\bibinfo {volume} {20}},\ \bibinfo {pages} {4176}
  (\bibinfo {year} {2012})}\BibitemShut {NoStop}%
\bibitem [{\citenamefont {David}\ and\ \citenamefont {{Garc\'ia de
  Abajo}}(2011)}]{david11}%
  \BibitemOpen
  \bibfield  {author} {\bibinfo {author} {\bibfnamefont {C.}~\bibnamefont
  {David}}\ and\ \bibinfo {author} {\bibfnamefont {F.~J.}\ \bibnamefont
  {{Garc\'ia de Abajo}}},\ }\href@noop {} {\bibfield  {journal} {\bibinfo
  {journal} {J. Phys. Chem. C}\ }\textbf {\bibinfo {volume} {115}},\ \bibinfo
  {pages} {19470} (\bibinfo {year} {2011})}\BibitemShut {NoStop}%
\bibitem [{\citenamefont {Toscano}\ \emph {et~al.}(2013)\citenamefont
  {Toscano}, \citenamefont {S.}, \citenamefont {Yan}, \citenamefont {Jeppesen},
  \citenamefont {Xiao}, \citenamefont {Wubs}, \citenamefont {Jauho},
  \citenamefont {Bozhevolnyi},\ and\ \citenamefont {Mortensen}}]{toscano13}%
  \BibitemOpen
  \bibfield  {author} {\bibinfo {author} {\bibfnamefont {G.}~\bibnamefont
  {Toscano}}, \bibinfo {author} {\bibfnamefont {R.}~\bibnamefont {S.}},
  \bibinfo {author} {\bibfnamefont {W.}~\bibnamefont {Yan}}, \bibinfo {author}
  {\bibfnamefont {C.}~\bibnamefont {Jeppesen}}, \bibinfo {author}
  {\bibfnamefont {S.}~\bibnamefont {Xiao}}, \bibinfo {author} {\bibfnamefont
  {M.}~\bibnamefont {Wubs}}, \bibinfo {author} {\bibfnamefont {A.-P.}\
  \bibnamefont {Jauho}}, \bibinfo {author} {\bibfnamefont {S.~I.}\ \bibnamefont
  {Bozhevolnyi}}, \ and\ \bibinfo {author} {\bibfnamefont {N.~A.}\ \bibnamefont
  {Mortensen}},\ }\href {\doibase 10.1515/nanoph-2013-0014} {\bibfield
  {journal} {\bibinfo  {journal} {Nanophotonics}\ }\textbf {\bibinfo {volume}
  {2}},\ \bibinfo {pages} {161–166} (\bibinfo {year} {2013})}\BibitemShut
  {NoStop}%
\bibitem [{\citenamefont {Yan}\ \emph {et~al.}(2013)\citenamefont {Yan},
  \citenamefont {Mortensen},\ and\ \citenamefont {Wubs}}]{yan13}%
  \BibitemOpen
  \bibfield  {author} {\bibinfo {author} {\bibfnamefont {W.}~\bibnamefont
  {Yan}}, \bibinfo {author} {\bibfnamefont {N.~A.}\ \bibnamefont {Mortensen}},
  \ and\ \bibinfo {author} {\bibfnamefont {M.}~\bibnamefont {Wubs}},\ }\href
  {\doibase 10.1103/PhysRevB.88.155414} {\bibfield  {journal} {\bibinfo
  {journal} {Phys. Rev. B}\ }\textbf {\bibinfo {volume} {88}},\ \bibinfo
  {pages} {155414} (\bibinfo {year} {2013})}\BibitemShut {NoStop}%
\bibitem [{\citenamefont {Mie}(1908)}]{mie08}%
  \BibitemOpen
  \bibfield  {author} {\bibinfo {author} {\bibfnamefont {G.}~\bibnamefont
  {Mie}},\ }\href@noop {} {\bibfield  {journal} {\bibinfo  {journal} {Ann.
  Phys.}\ }\textbf {\bibinfo {volume} {330}},\ \bibinfo {pages} {377} (\bibinfo
  {year} {1908})}\BibitemShut {NoStop}%
\bibitem [{\citenamefont {Ruppin}(1973)}]{ruppin73}%
  \BibitemOpen
  \bibfield  {author} {\bibinfo {author} {\bibfnamefont {R.}~\bibnamefont
  {Ruppin}},\ }\href@noop {} {\bibfield  {journal} {\bibinfo  {journal} {Phys.
  Rev. Lett.}\ }\textbf {\bibinfo {volume} {31}},\ \bibinfo {pages} {1434}
  (\bibinfo {year} {1973})}\BibitemShut {NoStop}%
\bibitem [{\citenamefont {Bohren}\ and\ \citenamefont
  {Huffman}(1983)}]{bohren}%
  \BibitemOpen
  \bibfield  {author} {\bibinfo {author} {\bibfnamefont {C.~F.}\ \bibnamefont
  {Bohren}}\ and\ \bibinfo {author} {\bibfnamefont {D.~R.}\ \bibnamefont
  {Huffman}},\ }\href@noop {} {\emph {\bibinfo {title} {Absorbtion and
  scattering of light by small particles}}},\ \bibinfo {edition} {1st}\ ed.\
  (\bibinfo  {publisher} {Wiley},\ \bibinfo {year} {1983})\BibitemShut
  {NoStop}%
\bibitem [{\citenamefont {Stradling}\ and\ \citenamefont
  {Wood}(1970)}]{stradling70}%
  \BibitemOpen
  \bibfield  {author} {\bibinfo {author} {\bibfnamefont {R.~A.}\ \bibnamefont
  {Stradling}}\ and\ \bibinfo {author} {\bibfnamefont {R.~A.}\ \bibnamefont
  {Wood}},\ }\href@noop {} {\bibfield  {journal} {\bibinfo  {journal} {J. Phys.
  C}\ }\textbf {\bibinfo {volume} {3}},\ \bibinfo {pages} {L94} (\bibinfo
  {year} {1970})}\BibitemShut {NoStop}%
\bibitem [{\citenamefont {Cunningham}\ and\ \citenamefont
  {Gruber}(1970)}]{cunningham70}%
  \BibitemOpen
  \bibfield  {author} {\bibinfo {author} {\bibfnamefont {R.~W.}\ \bibnamefont
  {Cunningham}}\ and\ \bibinfo {author} {\bibfnamefont {J.~B.}\ \bibnamefont
  {Gruber}},\ }\href@noop {} {\bibfield  {journal} {\bibinfo  {journal} {J.
  Appl. Phys}\ }\textbf {\bibinfo {volume} {41}},\ \bibinfo {pages} {1804}
  (\bibinfo {year} {1970})}\BibitemShut {NoStop}%
\bibitem [{\citenamefont {Szmyd}\ \emph {et~al.}(1990)\citenamefont {Szmyd},
  \citenamefont {Porro}, \citenamefont {Majerfeld},\ and\ \citenamefont
  {Lagomarsino}}]{szmyd90}%
  \BibitemOpen
  \bibfield  {author} {\bibinfo {author} {\bibfnamefont {D.~M.}\ \bibnamefont
  {Szmyd}}, \bibinfo {author} {\bibfnamefont {P.}~\bibnamefont {Porro}},
  \bibinfo {author} {\bibfnamefont {A.}~\bibnamefont {Majerfeld}}, \ and\
  \bibinfo {author} {\bibfnamefont {S.}~\bibnamefont {Lagomarsino}},\
  }\href@noop {} {\bibfield  {journal} {\bibinfo  {journal} {J. Appl. Phys.}\
  }\textbf {\bibinfo {volume} {68}},\ \bibinfo {pages} {2367} (\bibinfo {year}
  {1990})}\BibitemShut {NoStop}%
\bibitem [{\citenamefont {Walton}\ and\ \citenamefont
  {Mishra}(1968)}]{walton68}%
  \BibitemOpen
  \bibfield  {author} {\bibinfo {author} {\bibfnamefont {A.~K.}\ \bibnamefont
  {Walton}}\ and\ \bibinfo {author} {\bibfnamefont {U.~K.}\ \bibnamefont
  {Mishra}},\ }\href@noop {} {\bibfield  {journal} {\bibinfo  {journal} {J.
  Phys. C}\ }\textbf {\bibinfo {volume} {1}},\ \bibinfo {pages} {533} (\bibinfo
  {year} {1968})}\BibitemShut {NoStop}%
\bibitem [{\citenamefont {Rowell}(1988)}]{rowell88}%
  \BibitemOpen
  \bibfield  {author} {\bibinfo {author} {\bibfnamefont {N.~L.}\ \bibnamefont
  {Rowell}},\ }\href@noop {} {\bibfield  {journal} {\bibinfo  {journal}
  {Infrared Phys.}\ }\textbf {\bibinfo {volume} {28}},\ \bibinfo {pages} {37}
  (\bibinfo {year} {1988})}\BibitemShut {NoStop}%
\bibitem [{\citenamefont {Sze}\ and\ \citenamefont {Irvin}(1968)}]{sze67}%
  \BibitemOpen
  \bibfield  {author} {\bibinfo {author} {\bibfnamefont {S.~M.}\ \bibnamefont
  {Sze}}\ and\ \bibinfo {author} {\bibfnamefont {J.~C.}\ \bibnamefont
  {Irvin}},\ }\href@noop {} {\bibfield  {journal} {\bibinfo  {journal} {Solid
  Stat. Elec.}\ }\textbf {\bibinfo {volume} {11}},\ \bibinfo {pages} {599}
  (\bibinfo {year} {1968})}\BibitemShut {NoStop}%
\bibitem [{\citenamefont {Madelung}(2004)}]{madelung}%
  \BibitemOpen
  \bibfield  {author} {\bibinfo {author} {\bibfnamefont {O.}~\bibnamefont
  {Madelung}},\ }\href@noop {} {\emph {\bibinfo {title} {Semiconductors: Data
  handbook}}},\ \bibinfo {edition} {3rd}\ ed.\ (\bibinfo  {publisher}
  {Springer-Verlag Berlin},\ \bibinfo {year} {2004})\BibitemShut {NoStop}%
\bibitem [{\citenamefont {Tiggesbäumker}\ \emph {et~al.}(1993)\citenamefont
  {Tiggesbäumker}, \citenamefont {Köller}, \citenamefont {Meiwes-Broer},\
  and\ \citenamefont {Liebsch}}]{tiggesbaumker93}%
  \BibitemOpen
  \bibfield  {author} {\bibinfo {author} {\bibfnamefont {J.}~\bibnamefont
  {Tiggesbäumker}}, \bibinfo {author} {\bibfnamefont {L.}~\bibnamefont
  {Köller}}, \bibinfo {author} {\bibfnamefont {K.-H.}\ \bibnamefont
  {Meiwes-Broer}}, \ and\ \bibinfo {author} {\bibfnamefont {A.}~\bibnamefont
  {Liebsch}},\ }\href@noop {} {\bibfield  {journal} {\bibinfo  {journal} {Phys.
  Rev. A}\ }\textbf {\bibinfo {volume} {48}},\ \bibinfo {pages} {R1749}
  (\bibinfo {year} {1993})}\BibitemShut {NoStop}%
\bibitem [{\citenamefont {Lindau}\ and\ \citenamefont
  {Nilsson}(1971)}]{lindau71}%
  \BibitemOpen
  \bibfield  {author} {\bibinfo {author} {\bibfnamefont {I.}~\bibnamefont
  {Lindau}}\ and\ \bibinfo {author} {\bibfnamefont {P.~O.}\ \bibnamefont
  {Nilsson}},\ }\href@noop {} {\bibfield  {journal} {\bibinfo  {journal} {Phys.
  Scr.}\ }\textbf {\bibinfo {volume} {3}},\ \bibinfo {pages} {87} (\bibinfo
  {year} {1971})}\BibitemShut {NoStop}%
\bibitem [{\citenamefont {Billaud}\ \emph {et~al.}(2010)\citenamefont
  {Billaud}, \citenamefont {Marhaba}, \citenamefont {Grillet}, \citenamefont
  {Cottancin}, \citenamefont {Bonnet}, \citenamefont {Lermé}, \citenamefont
  {Vialle}, \citenamefont {Broyer},\ and\ \citenamefont
  {Pellarin}}]{billaud10}%
  \BibitemOpen
  \bibfield  {author} {\bibinfo {author} {\bibfnamefont {P.}~\bibnamefont
  {Billaud}}, \bibinfo {author} {\bibfnamefont {S.}~\bibnamefont {Marhaba}},
  \bibinfo {author} {\bibfnamefont {N.}~\bibnamefont {Grillet}}, \bibinfo
  {author} {\bibfnamefont {E.}~\bibnamefont {Cottancin}}, \bibinfo {author}
  {\bibfnamefont {C.}~\bibnamefont {Bonnet}}, \bibinfo {author} {\bibfnamefont
  {J.}~\bibnamefont {Lermé}}, \bibinfo {author} {\bibfnamefont {J.-L.}\
  \bibnamefont {Vialle}}, \bibinfo {author} {\bibfnamefont {M.}~\bibnamefont
  {Broyer}}, \ and\ \bibinfo {author} {\bibfnamefont {M.}~\bibnamefont
  {Pellarin}},\ }\href@noop {} {\bibfield  {journal} {\bibinfo  {journal} {Rev.
  Sci. Instr.}\ }\textbf {\bibinfo {volume} {81}},\ \bibinfo {pages} {043101}
  (\bibinfo {year} {2010})}\BibitemShut {NoStop}%
\bibitem [{\citenamefont {Howells}\ and\ \citenamefont
  {Schlie}(1996)}]{howells96}%
  \BibitemOpen
  \bibfield  {author} {\bibinfo {author} {\bibfnamefont {S.~C.}\ \bibnamefont
  {Howells}}\ and\ \bibinfo {author} {\bibfnamefont {L.~A.}\ \bibnamefont
  {Schlie}},\ }\href@noop {} {\bibfield  {journal} {\bibinfo  {journal} {Appl.
  Phys. Lett.}\ }\textbf {\bibinfo {volume} {69}},\ \bibinfo {pages} {550}
  (\bibinfo {year} {1996})}\BibitemShut {NoStop}%
\bibitem [{\citenamefont {Charlé}\ \emph {et~al.}(1998)\citenamefont
  {Charlé}, \citenamefont {König}, \citenamefont {Nepijko}, \citenamefont
  {Rabin},\ and\ \citenamefont {Schulze}}]{charle98}%
  \BibitemOpen
  \bibfield  {author} {\bibinfo {author} {\bibfnamefont {K.-P.}\ \bibnamefont
  {Charlé}}, \bibinfo {author} {\bibfnamefont {L.}~\bibnamefont {König}},
  \bibinfo {author} {\bibfnamefont {S.}~\bibnamefont {Nepijko}}, \bibinfo
  {author} {\bibfnamefont {I.}~\bibnamefont {Rabin}}, \ and\ \bibinfo {author}
  {\bibfnamefont {W.}~\bibnamefont {Schulze}},\ }\href@noop {} {\bibfield
  {journal} {\bibinfo  {journal} {Cryst. Res. Technol.}\ }\textbf {\bibinfo
  {volume} {33}},\ \bibinfo {pages} {1085} (\bibinfo {year}
  {1998})}\BibitemShut {NoStop}%
\bibitem [{\citenamefont {Raza}\ \emph
  {et~al.}(2013{\natexlab{c}})\citenamefont {Raza}, \citenamefont {Stenger},
  \citenamefont {Kadkhodazadeh}, \citenamefont {Fischer}, \citenamefont
  {Kostesha}, \citenamefont {Jauho}, \citenamefont {Burrows}, \citenamefont
  {Wubs},\ and\ \citenamefont {A.}}]{raza13b}%
  \BibitemOpen
  \bibfield  {author} {\bibinfo {author} {\bibfnamefont {S.}~\bibnamefont
  {Raza}}, \bibinfo {author} {\bibfnamefont {N.}~\bibnamefont {Stenger}},
  \bibinfo {author} {\bibfnamefont {S.}~\bibnamefont {Kadkhodazadeh}}, \bibinfo
  {author} {\bibfnamefont {S.~V.}\ \bibnamefont {Fischer}}, \bibinfo {author}
  {\bibfnamefont {N.}~\bibnamefont {Kostesha}}, \bibinfo {author}
  {\bibfnamefont {A.-P.}\ \bibnamefont {Jauho}}, \bibinfo {author}
  {\bibfnamefont {A.}~\bibnamefont {Burrows}}, \bibinfo {author} {\bibfnamefont
  {M.}~\bibnamefont {Wubs}}, \ and\ \bibinfo {author} {\bibfnamefont {M.~N.}\
  \bibnamefont {A.}},\ }\href@noop {} {\bibfield  {journal} {\bibinfo
  {journal} {Nanophotonics}\ }\textbf {\bibinfo {volume} {2}},\ \bibinfo
  {pages} {131} (\bibinfo {year} {2013}{\natexlab{c}})}\BibitemShut {NoStop}%
\bibitem [{\citenamefont {Raza}\ \emph
  {et~al.}(2015{\natexlab{b}})\citenamefont {Raza}, \citenamefont
  {Kadkhodazadeh}, \citenamefont {Christensen}, \citenamefont {{Di Vece}},
  \citenamefont {Wubs}, \citenamefont {Mortensen},\ and\ \citenamefont
  {Stenger}}]{raza15b}%
  \BibitemOpen
  \bibfield  {author} {\bibinfo {author} {\bibfnamefont {S.}~\bibnamefont
  {Raza}}, \bibinfo {author} {\bibfnamefont {S.}~\bibnamefont {Kadkhodazadeh}},
  \bibinfo {author} {\bibfnamefont {T.}~\bibnamefont {Christensen}}, \bibinfo
  {author} {\bibfnamefont {M.}~\bibnamefont {{Di Vece}}}, \bibinfo {author}
  {\bibfnamefont {M.}~\bibnamefont {Wubs}}, \bibinfo {author} {\bibfnamefont
  {N.~A.}\ \bibnamefont {Mortensen}}, \ and\ \bibinfo {author} {\bibfnamefont
  {N.}~\bibnamefont {Stenger}},\ }\href {\doibase 10.1038/ncomms9788}
  {\bibfield  {journal} {\bibinfo  {journal} {Nat. Commun.}\ }\textbf {\bibinfo
  {volume} {6}},\ \bibinfo {pages} {8788} (\bibinfo {year}
  {2015}{\natexlab{b}})}\BibitemShut {NoStop}%
\bibitem [{\citenamefont {Scholl}\ \emph {et~al.}(2012)\citenamefont {Scholl},
  \citenamefont {Koh},\ and\ \citenamefont {Dionne}}]{Scholl12}%
  \BibitemOpen
  \bibfield  {author} {\bibinfo {author} {\bibfnamefont {J.~A.}\ \bibnamefont
  {Scholl}}, \bibinfo {author} {\bibfnamefont {A.~L.}\ \bibnamefont {Koh}}, \
  and\ \bibinfo {author} {\bibfnamefont {J.~A.}\ \bibnamefont {Dionne}},\
  }\href {\doibase 10.1038/nature10904} {\bibfield  {journal} {\bibinfo
  {journal} {Nature}\ }\textbf {\bibinfo {volume} {483}},\ \bibinfo {pages}
  {421} (\bibinfo {year} {2012})}\BibitemShut {NoStop}%
\bibitem [{\citenamefont {Yan}\ and\ \citenamefont {Mortensen}(2016)}]{yan16}%
  \BibitemOpen
  \bibfield  {author} {\bibinfo {author} {\bibfnamefont {W.}~\bibnamefont
  {Yan}}\ and\ \bibinfo {author} {\bibfnamefont {N.~A.}\ \bibnamefont
  {Mortensen}},\ }\href {\doibase 10.1103/PhysRevB.93.115439} {\bibfield
  {journal} {\bibinfo  {journal} {Phys. Rev. B}\ }\textbf {\bibinfo {volume}
  {93}},\ \bibinfo {pages} {115439} (\bibinfo {year} {2016})}\BibitemShut
  {NoStop}%
\bibitem [{\citenamefont {Werner}\ \emph {et~al.}(2009)\citenamefont {Werner},
  \citenamefont {Glantschnig},\ and\ \citenamefont
  {Ambrosch-Draxl}}]{werner09}%
  \BibitemOpen
  \bibfield  {author} {\bibinfo {author} {\bibfnamefont {W.~S.~M.}\
  \bibnamefont {Werner}}, \bibinfo {author} {\bibfnamefont {K.}~\bibnamefont
  {Glantschnig}}, \ and\ \bibinfo {author} {\bibfnamefont {C.}~\bibnamefont
  {Ambrosch-Draxl}},\ }\href@noop {} {\bibfield  {journal} {\bibinfo  {journal}
  {J. Phys. Chem. Ref. Data}\ }\textbf {\bibinfo {volume} {38}},\ \bibinfo
  {pages} {1013} (\bibinfo {year} {2009})}\BibitemShut {NoStop}%
\bibitem [{\citenamefont {Gaur}(1976)}]{gaur76}%
  \BibitemOpen
  \bibfield  {author} {\bibinfo {author} {\bibfnamefont {N.~K.~S.}\
  \bibnamefont {Gaur}},\ }\href@noop {} {\bibfield  {journal} {\bibinfo
  {journal} {Physica B \& C}\ }\textbf {\bibinfo {volume} {82}},\ \bibinfo
  {pages} {262} (\bibinfo {year} {1976})}\BibitemShut {NoStop}%
\bibitem [{\citenamefont {Gu}\ \emph {et~al.}(2000)\citenamefont {Gu},
  \citenamefont {Tani}, \citenamefont {Sakai},\ and\ \citenamefont
  {Yang}}]{gu00}%
  \BibitemOpen
  \bibfield  {author} {\bibinfo {author} {\bibfnamefont {P.}~\bibnamefont
  {Gu}}, \bibinfo {author} {\bibfnamefont {M.}~\bibnamefont {Tani}}, \bibinfo
  {author} {\bibfnamefont {K.}~\bibnamefont {Sakai}}, \ and\ \bibinfo {author}
  {\bibfnamefont {T.-R.}\ \bibnamefont {Yang}},\ }\href@noop {} {\bibfield
  {journal} {\bibinfo  {journal} {Appl. Phys. Lett.}\ }\textbf {\bibinfo
  {volume} {77}},\ \bibinfo {pages} {1798} (\bibinfo {year}
  {2000})}\BibitemShut {NoStop}%
\bibitem [{\citenamefont {Olsen}\ and\ \citenamefont {Lynch}(1969)}]{olsen69}%
  \BibitemOpen
  \bibfield  {author} {\bibinfo {author} {\bibfnamefont {C.~G.}\ \bibnamefont
  {Olsen}}\ and\ \bibinfo {author} {\bibfnamefont {D.~W.}\ \bibnamefont
  {Lynch}},\ }\href@noop {} {\bibfield  {journal} {\bibinfo  {journal} {Phys.
  Rev.}\ }\textbf {\bibinfo {volume} {177}},\ \bibinfo {pages} {1231} (\bibinfo
  {year} {1969})}\BibitemShut {NoStop}%
\bibitem [{\citenamefont {Ritz}\ and\ \citenamefont {Lüth}(1985)}]{ritz85}%
  \BibitemOpen
  \bibfield  {author} {\bibinfo {author} {\bibfnamefont {A.}~\bibnamefont
  {Ritz}}\ and\ \bibinfo {author} {\bibfnamefont {H.}~\bibnamefont {Lüth}},\
  }\href@noop {} {\bibfield  {journal} {\bibinfo  {journal} {J. Vac. Sci.
  Technol. B}\ }\textbf {\bibinfo {volume} {3}},\ \bibinfo {pages} {1153}
  (\bibinfo {year} {1985})}\BibitemShut {NoStop}%
\bibitem [{\citenamefont {Bell}\ \emph {et~al.}(1998)\citenamefont {Bell},
  \citenamefont {Jones},\ and\ \citenamefont {McConville}}]{bell98}%
  \BibitemOpen
  \bibfield  {author} {\bibinfo {author} {\bibfnamefont {G.~R.}\ \bibnamefont
  {Bell}}, \bibinfo {author} {\bibfnamefont {T.~S.}\ \bibnamefont {Jones}}, \
  and\ \bibinfo {author} {\bibfnamefont {C.~F.}\ \bibnamefont {McConville}},\
  }\href@noop {} {\bibfield  {journal} {\bibinfo  {journal} {Surf. Sci.}\
  }\textbf {\bibinfo {volume} {405}},\ \bibinfo {pages} {280} (\bibinfo {year}
  {1998})}\BibitemShut {NoStop}%
\bibitem [{\citenamefont {Adomavicius}\ \emph {et~al.}(2009)\citenamefont
  {Adomavicius}, \citenamefont {Macutkevic}, \citenamefont {Suzanoviciene},
  \citenamefont {Šiušys},\ and\ \citenamefont {Krotkus}}]{adomavicius09}%
  \BibitemOpen
  \bibfield  {author} {\bibinfo {author} {\bibfnamefont {R.}~\bibnamefont
  {Adomavicius}}, \bibinfo {author} {\bibfnamefont {J.}~\bibnamefont
  {Macutkevic}}, \bibinfo {author} {\bibfnamefont {R.}~\bibnamefont
  {Suzanoviciene}}, \bibinfo {author} {\bibfnamefont {A.}~\bibnamefont
  {Šiušys}}, \ and\ \bibinfo {author} {\bibfnamefont {A.}~\bibnamefont
  {Krotkus}},\ }\href@noop {} {\bibfield  {journal} {\bibinfo  {journal} {Phys.
  Status Solidi C}\ }\textbf {\bibinfo {volume} {6}},\ \bibinfo {pages} {2849}
  (\bibinfo {year} {2009})}\BibitemShut {NoStop}%
\bibitem [{\citenamefont {Christensen}\ \emph
  {et~al.}(2014{\natexlab{b}})\citenamefont {Christensen}, \citenamefont
  {Wang}, \citenamefont {Jauho}, \citenamefont {Wubs},\ and\ \citenamefont
  {Mortensen}}]{christensen14b}%
  \BibitemOpen
  \bibfield  {author} {\bibinfo {author} {\bibfnamefont {T.}~\bibnamefont
  {Christensen}}, \bibinfo {author} {\bibfnamefont {W.}~\bibnamefont {Wang}},
  \bibinfo {author} {\bibfnamefont {A.-P.}\ \bibnamefont {Jauho}}, \bibinfo
  {author} {\bibfnamefont {M.}~\bibnamefont {Wubs}}, \ and\ \bibinfo {author}
  {\bibfnamefont {N.~A.}\ \bibnamefont {Mortensen}},\ }\href@noop {} {\bibfield
   {journal} {\bibinfo  {journal} {Phys. Rev. B}\ }\textbf {\bibinfo {volume}
  {90}},\ \bibinfo {pages} {241414} (\bibinfo {year}
  {2014}{\natexlab{b}})}\BibitemShut {NoStop}%
\bibitem [{\citenamefont {Lundeberg}\ \emph {et~al.}()\citenamefont
  {Lundeberg}, \citenamefont {Gao}, \citenamefont {Asgari}, \citenamefont
  {Tan}, \citenamefont {Duppen}, \citenamefont {Autore}, \citenamefont
  {Alonso-González}, \citenamefont {Woessner}, \citenamefont {Watanabe},
  \citenamefont {Taniguchi}, \citenamefont {Hillenbrand}, \citenamefont {Hone},
  \citenamefont {Polini},\ and\ \citenamefont {Koppens}}]{lundberg17}%
  \BibitemOpen
  \bibfield  {author} {\bibinfo {author} {\bibfnamefont {M.~B.}\ \bibnamefont
  {Lundeberg}}, \bibinfo {author} {\bibfnamefont {Y.}~\bibnamefont {Gao}},
  \bibinfo {author} {\bibfnamefont {R.}~\bibnamefont {Asgari}}, \bibinfo
  {author} {\bibfnamefont {C.}~\bibnamefont {Tan}}, \bibinfo {author}
  {\bibfnamefont {B.~V.}\ \bibnamefont {Duppen}}, \bibinfo {author}
  {\bibfnamefont {M.}~\bibnamefont {Autore}}, \bibinfo {author} {\bibfnamefont
  {P.}~\bibnamefont {Alonso-González}}, \bibinfo {author} {\bibfnamefont
  {A.}~\bibnamefont {Woessner}}, \bibinfo {author} {\bibfnamefont
  {K.}~\bibnamefont {Watanabe}}, \bibinfo {author} {\bibfnamefont
  {T.}~\bibnamefont {Taniguchi}}, \bibinfo {author} {\bibfnamefont
  {R.}~\bibnamefont {Hillenbrand}}, \bibinfo {author} {\bibfnamefont
  {J.}~\bibnamefont {Hone}}, \bibinfo {author} {\bibfnamefont {M.}~\bibnamefont
  {Polini}}, \ and\ \bibinfo {author} {\bibfnamefont {F.~H.~L.}\ \bibnamefont
  {Koppens}},\ }\href@noop {} {\enquote {\bibinfo {title} {Tuning quantum
  non-local effects in graphene plasmonics},}\ }\bibinfo {note} {To be
  published in \emph{Science} (2017), DOI: 10.1126/science.aan2735}\BibitemShut
  {NoStop}%
\bibitem [{\citenamefont {Bell}\ \emph {et~al.}(2006)\citenamefont {Bell},
  \citenamefont {Veal}, \citenamefont {Frost},\ and\ \citenamefont
  {McConville}}]{bell06}%
  \BibitemOpen
  \bibfield  {author} {\bibinfo {author} {\bibfnamefont {G.~R.}\ \bibnamefont
  {Bell}}, \bibinfo {author} {\bibfnamefont {T.~D.}\ \bibnamefont {Veal}},
  \bibinfo {author} {\bibfnamefont {J.~A.}\ \bibnamefont {Frost}}, \ and\
  \bibinfo {author} {\bibfnamefont {C.~F.}\ \bibnamefont {McConville}},\
  }\href@noop {} {\bibfield  {journal} {\bibinfo  {journal} {Phys. Rev. B}\
  }\textbf {\bibinfo {volume} {73}},\ \bibinfo {pages} {153302} (\bibinfo
  {year} {2006})}\BibitemShut {NoStop}%
\end{thebibliography}%

\end{document}